\documentclass[aps,pra,twocolumn,superscriptaddress,amsmath,amssymb,amsfonts,nofootinbib]{revtex4-1}
\usepackage{graphicx}
\usepackage{longtable}
\usepackage{color}

\newcommand{\commut}[2]{\left[ #1 , #2 \right]}

\newcommand{\ket}[1]{\vert #1 \rangle}
\newcommand{\bra}[1]{\langle #1 \vert}

\newcommand{\proj}[1]{\ket{#1}\bra{#1}}

\newcommand{\be}{\begin{equation}}
\newcommand{\ee}{\end{equation}}
\newcommand{\id}{\mathbb{I}}

\newcommand{\ba}{\begin{align}}
\newcommand{\ea}{\end{align}}
\newcommand{\bas}{\begin{align*}}
\newcommand{\eas}{\end{align*}}
\newcommand{\barr}[1]{\begin{array}{#1}}
\newcommand{\earr}{\end{array}}

\newcommand{\matdeux}[1]{\left( \barr{cc} #1 \earr \right)}

\newcommand{\matquatre}[1]{\left( \barr{cccc} #1 \earr \right)}

\newcommand{\eul}[1]{\mathrm{e}^{#1}}
\newcommand{\sx}{\sigma_x}	

\newcommand{\sz}{\sigma_z}

\newcommand{\sm}{\sigma_{-}}

\newcommand{\mrm}[1]{\mathrm{#1}}

\newcommand{\ad}{a^\dagger}
\newcommand{\bd}{b^\dagger}

\newcommand{\eq}[1]{(\ref{#1})}

\begin{document}

\title{First order sidebands in circuit QED using qubit frequency modulation}

\author{F\'elix Beaudoin}
\email[]{felix.beaudoin@mail.mcgill.ca}
\affiliation{D\'epartement de Physique, Universit\'e de Sherbrooke, Sherbrooke, Qu\'ebec, Canada J1K 2R1}
\affiliation{Quantum Information Processing Group, Raytheon BBN Technologies, Cambridge, MA 02138, USA}
\affiliation{Department of Physics, McGill University, Montr\'eal, Qu\'ebec, Canada H3A 2T8}
\author{Marcus P. da Silva}
\affiliation{Quantum Information Processing Group, Raytheon BBN Technologies, Cambridge, MA 02138, USA}
\author{Zachary Dutton}
\affiliation{Quantum Information Processing Group, Raytheon BBN Technologies, Cambridge, MA 02138, USA}
\author{Alexandre Blais}
\affiliation{D\'epartement de Physique, Universit\'e de Sherbrooke, Sherbrooke, Qu\'ebec, Canada J1K 2R1}

\date{\today}

\begin{abstract}
	Sideband transitions have been shown to generate controllable interaction between superconducting qubits and microwave resonators. Up to now, these transitions have been implemented with voltage drives on the qubit or the resonator, with the significant disadvantage that such implementations only lead to second-order sideband transitions. Here we propose an approach to achieve first-order sideband transitions by relying on controlled oscillations of the qubit frequency using a flux-bias line.  Not only can first-order transitions be significantly faster, but the same technique can be employed to implement other tunable qubit-resonator and qubit-qubit interactions.  We discuss in detail how such first-order sideband transitions can be used to implement a high fidelity CNOT operation between two transmons coupled to the same resonator.
\end{abstract}

\pacs{03.67.Lx, 03.67.Bg, 42.50.Pq, 85.25.Cp}

\maketitle

\section{Introduction}

In circuit QED, superconducting qubits are coupled to a common mode of a transmission line resonator~\cite{blais2004cavity,wallraff2004strong} or 3D cavity~\cite{PhysRevLett.107.240501}. The shared mode can be used to mediate interaction between the otherwise non-interacting qubits. Several mechanisms to realize two-qubit gates in this system have been proposed~\cite{blais2007quantum,PhysRevB.78.064503,PhysRevB.81.134507,PhysRevB.82.024514} and implemented~\cite{dicarlo2009demonstration,majer2007coupling,Mariantoni07102011,fedorov2011implementation,reed2012realization}. One approach relies on sideband transitions~\cite{blais2007quantum} where, for example, the qubit state can be swapped with the resonator state. By combining sideband transitions involving two qubits, an entangling two-qubit gate can be implemented~\cite{cirac1995quantum,PhysRevB.79.180511}.

In practice, sideband transitions are realized by voltage-driving the resonator~\cite{blais2007quantum,wallraff2007sideband} or the qubit~\cite{PhysRevB.79.180511,PhysRevLett.104.100504}. Because this type of drive can only cause transitions between states of different parity, and since the two states involved in any sideband transition have the same parity, voltage driving can only be used in a second order process~\cite{blais2007quantum}. In other words, in circuit QED, all sideband transitions studied so far have relied on two-photon processes, leading to low transition rates and consequently slow two-qubit gates.

In this paper, we propose to speed up gates based on sideband transtions by relying on a different kind of drive. Indeed, in many current experiments~\cite{hofheinz2008generation,hofheinz2009synthesizing,dicarlo2009demonstration,dicarlo2010preparation}, the qubit transition frequency can be adjusted by several hundreds of MHz in a few nanoseconds. This is made possible by a fast-flux line fabricated in close proximity to the qubit's loop and is described by the Hamiltonian
\be	\label{eqn:fm}
	H_\mrm{FC}=\frac{f(t)}{2}\sz.
\ee
Here, $f(t)$ represents the qubit frequency change and is a function of the externally applied flux. By modulating the function $f(t)\propto \cos(\omega_\mathrm{FC} t)$, and choosing the modulation frequency $\omega_\mathrm{FC}$ appropriately, we show that sideband transitions can be generated. Since this frequency control (FC) signal is proportional to $\sz$ (and not $\sx$ as is the case for the more typical voltage driving), sidebands can be generated in a first order process and consequently result in faster two-qubit gates. Moreover, in the presence of two-qubits coupled to the same mode of a resonator, this frequency modulation can result in a second-order flip-flop interaction between the qubits in a way similar to the FLICFORQ~\cite{rigetti2005protocol} and the cross-resonance~\cite{PhysRevLett.107.080502} gates.

This paper is divided as follows. In Section~\ref{sec:tunable}, we give the Hamiltonian for two qubits coupled to the same mode of a resonator, including an FC drive on both qubits. By moving to the frame that diagonalizes the Hamiltonian, we find that the FC drive generates multiple effective qubit-resonator and qubit-qubit interactions. These interactions can be essentially turned on and off by choosing the appropriate modulation frequency $\omega_\mathrm{FC}$. Here, we focus on sideband transitions which are shown to be of first order in the modulation amplitude. In Section~\ref{sec:SB}, we calculate the dynamics associated to the red sideband process between a qubit and the resonator, and consider the error coming from the interaction with a spectator qubit. These ideas are then extended in Section~\ref{sec:mls} to the more realistic case of many-level systems (MLSs) in a resonator. We also show that the presence of a second excited state can be exploited in a pulse sequence containing red sidebands to generate a two-qubit gate which is locally equivalent to a CNOT. Finally, in Section~\ref{sec:transmon}, we explain how our proposal can be implemented with transmon qubits. We first consider the case of a single red sideband and show how possible caveats can be avoided. At last, we present simulation results that predict that FC modulations can be used to generate high-fidelity maximally entangling gates with two transmons.

\section{Effective tunable interactions}
\label{sec:tunable}

In this section, we first present the lab frame Hamiltonian for two qubits coupled to a single mode of a resonator. Assuming the qubits to be well detuned from the resonator and ignoring the presence of any driving terms for the moment, we eliminate the resonant qubit-resonator interaction using a dispersive transformation. This results in, in particular, a residual qubit-qubit virtual interaction which we diagonalize. Reintroducing the FC drive Hamiltonians in that diagonal frame, we obtain the expressions for the tunable interactions, among which the red and blue sideband transitions are present. To simplify the discussion, we will ignore here the presence of higher excited states of the qubits. These levels are important,  for example, in the transmon qubit and their effect will be discussed in Section~\ref{sec:mls}.

\subsection{Lab frame Hamiltonian and generalized dispersive Hamiltonian}
\label{sec:gen:disp}

The electric-dipole coupling of two qubits to a single mode of a resonator is described by the Rabi Hamiltonian, with $\hbar=1$,~\cite{blais2004cavity}
\begin{align}
	H=\omega_r\ad a+\sum_{k=1,2} \left[\frac{\omega_a^{(k)}}{2}\sz^k+g^{(k)}X\sx^k\right].
\end{align}
Here, $\omega_r$ is the resonator frequency, $\omega_a^{(k)}$ the transition frequency of qubit $k$, $g^{(k)}$ the qubit-resonator coupling strength for qubit $k$ and $X=\ad+a$. The free qubit eigenstates will be labelled $\{\ket g,\;\ket e\}$. 

To approximately diagonalize $H$ we use the generalized dispersive transformation based on the unitary~\cite{hausinger2008dissipative,PhysRevA.84.043832}
\be
	U_\mrm  D = \exp\!\!\left[\!\sum_{k=1,2}\!\!\lambda^{(k)}\ad\sm^k\!+\!\Lambda^{(k)} a\sm^k\!+\!\xi^{(k)}\sz^k a^2\!-\!\mrm{H.c.}\!\right]\!\!,\!\!\label{eqn:tls:transform}
\ee
where $\lambda^{(k)}=g^{(k)}/\Delta^{(k)}$, $\Lambda^{(k)}=g^{(k)}/\Sigma^{(k)}$ with $\Delta^{(k)}=\omega_a^{(k)}-\omega_r$ and $\Sigma^{(k)}=\omega_a^{(k)}+\omega_r$. 
The squeezing-like terms are proportional to $\xi^{(k)}=(\chi^{(k)}+\mu^{(k)})/4\omega_r$, where $\chi^{(k)}=g^{(k)}\lambda^{(k)}$ and $\mu^{(k)}=g^{(k)}\Lambda{(k)}$ are respectively the dispersive and Bloch-Siegert shifts. Using the Campbell-Baker-Hausdorff relation
\be\label{eqn-campbell}
	\eul{-X}H\eul{X}= H+\commut{H}{X}+\frac{1}{2!}\commut{\commut{H}{X}}{X}+\ldots
\ee
we obtain, to second order in the couplings $g^{(k)}$,
\begin{equation}\label{eqn:J:Hamiltonian}
\begin{split}
	H_\mrm D 
	& = U_\mrm D^\dagger H U_\mrm D \\
	& \simeq \omega_r\ad a+\sum_{k=1,2}\frac{\tilde\omega_a^{(k)}}{2}\sz^k+S^{(k)}\sz^k\ad a 
		+J\sx^1\sx^2,
\end{split}
\end{equation}
where $\tilde\omega_a^{(k)}=\omega_a^{(k)}+\chi^{(k)}+\mu^{(k)}$ are the Lamb-shifted qubit frequencies and $S^{(k)}=\chi^{(k)}+\mu^{(k)}$ is the ac-Stark shift per photon. The last term of $H_\mrm D$, proportional to 
\be
	J=\frac{g^{(1)}g^{(2)}}{2}\sum_{k=1,2}\left(\frac{1}{\Delta^{(k)}}-\frac{1}{\Sigma^{(k)}}\right),
\ee
is a qubit-qubit interaction mediated by the exchange of virtual photons.

\subsection{Diagonal Hamiltonian}

Because of this last term, $H_\mrm D$ is not diagonal. It however has a simple structure which we now exploit. Indeed, since $H_\mrm D$ is diagonal with respect to the cavity degree of freedom, it is useful to focus on subspaces labelled by the fixed photon number $n$. In these subspaces, the Hamiltonian effectively reduces to a $4\times4$ matrix
\begin{align}
	H_\mrm D^n=\matquatre{
			\tilde\Sigma_\mrm Q^n/2 & 0 & 0 & J\\
			0 & \tilde\Delta_\mrm Q^n/2 & J & 0\\
			0 & J & -\tilde\Delta_\mrm Q^n/2 & 0\\
			J & 0 & 0 & -\tilde\Sigma_\mrm Q^n/2
		},\label{eqn:blocks}
\end{align}
with
\begin{align}
	\tilde\Delta_\mrm Q^n&=\Delta_\mrm Q+2n\Delta_\mrm S\\
	\tilde\Sigma_\mrm Q^n&=\Sigma_\mrm Q+2n\Sigma_\mrm S
\end{align}
and where we have defined $\Delta_\mrm Q = \tilde\omega_a^{(1)}-\tilde\omega_a^{(2)}$, $\Delta_\mrm S=S^{(1)}-S^{(2)}$, $\Sigma_\mrm Q = \tilde\omega_a^{(1)}+\tilde\omega_a^{(2)}$, and $\Sigma_\mrm S=S^{(1)}+S^{(2)}$. Because of its block structure, this Hamiltonian can easily be diagonalized. As in the cross-resonance gate~\cite{PhysRevB.81.134507,PhysRevLett.107.080502}, we will define two effective (or logical) qubits out of the eigenstates of this 4-dimensional block, the dependence on $n$ simply leading to ac-Stark shifts of these effective qubits. As discussed in details in Appendix~\ref{sec:diag}, if the two qubits are far-detuned from each other, the resulting diagonal Hamiltonian is, to second order in $J/\Delta_\mrm Q$ and $J/\Sigma_\mrm Q$, given by
\begin{align}
	H_\mrm{diag}=\omega_r\ad a&+\left(\tilde{\omega}_a^{(1)}+2\ad a S^{(1)}+S_\mrm J^+\right)\frac{\tau_z^1}{2}\notag\\
		&+\left(\tilde{\omega}_a^{(2)}+2\ad a S^{(2)}+S_\mrm J^-\right)\frac{\tau_z^2}{2}\label{eq:Hdiag:disp},
\end{align}
where
\be\label{eq:Spm}
	S_\mrm J^\pm=J^2\left(\frac{1}{\Sigma_\mrm Q}\pm\frac{1}{\Delta_\mrm Q}\right).
\ee
Here, we have also defined the Pauli operators $\tau_z^k$ acting on the effective qubits. In this diagonal frame, the interacting physical qubits are replaced by two logical non-interacting (and non-local) qubits, whose eigenstates will be denoted $\{\ket{0},\ket{1}\}$. The effect of the resonator-mediated $J$ interaction between the physical qubits is to shift the logical qubits by a quantity $S_J^\pm$ which is assumed here to be small (see Appendix~\ref{sec:diag} for the general result). Since it is of fourth order in $g^{(k)}$, this shift can be neglected, as will be done in the rest of the paper.

\subsection{FC drive in the diagonal frame}

We now introduce an FC drive on each qubit. In the lab frame, this takes the form
\be
	H_\mrm{FC}(t)=\frac{f^{(1)}(t)}{2}\sz^1+\frac{f^{(2)}(t)}{2}\sz^2~\label{eqn:fm:2}.
\ee
Later on, we will take
\be
	f^{(k)}(t) = \varepsilon^{(k)}(t)\cos\left[\omega_\mrm{FC}^{(k)}t\right],\label{eq:fk}
\ee
where $\varepsilon^{(k)}(t)$ is an envelope that varies slowly compared to the modulation frequency $\omega_\mrm{FC}^{(k)}$. Going to the dispersive frame and then to the diagonal frame as above, this Hamiltonian becomes
\begin{align}
	H_\mrm{FC}^\mrm{diag}(t)=&\;H_z^1(t)+H_z^2(t)+H_\mrm{SB}^1(t)+H_\mrm{SB}^2(t)\notag\\
		&+H_\mrm{PO}(t)+H_\mrm{QQ}(t)\label{eq:FC:H}.
\end{align}
The first two terms represent modulation in the effective qubit splittings
\begin{align}
	H_z^1(t)\simeq&\left[(1-\hat\lambda_\mrm J^2)\hat s_n^{(1)}f^{(1)}(t)+\hat\lambda_\mrm J^2\hat s_n^{(2)}f^{(2)}(t)\right]\frac{\tau_z^1}{2},\\
	H_z^2(t)\simeq&\left[\hat\lambda_\mrm J^2\hat s_n^{(1)}f^{(1)}(t)+(1-\hat\lambda_\mrm J^2)\hat s_n^{(2)}f^{(2)}(t)\right]\frac{\tau_z^2}{2},
\end{align}
where we have defined
\begin{align}
	\hat s_n^{(k)}&=1-\left[\left(\lambda^{(k)}\right)^2+\left(\Lambda^{(k)}\right)^2\right]\left(\ad a+1/2\right),\\
	\hat \lambda_\mrm J&=\frac{J}{\Delta_\mrm Q+2\ad a\Delta_\mrm S}.
\end{align}
As will be seen in Section~\ref{sec:FC:transmon}, these terms can lead to local phase gates on the logical qubit. Just like in the single qubit case~\cite{PhysRevA.84.043832}, in addition to these frequency modulations, we also get sideband transitions and parametric oscillations. However, these processes are modified by the qubit-qubit interactions. First, we find
\begin{align}
	H_\mrm{SB}^1(t)\simeq&-f^{(1)}(t)\Big[\lambda^{(1)}\ad\tau_-^1+\Lambda^{(1)}\ad\tau_+^1\\
		&+\hat\lambda_\mrm J \tau_z^1\left(\lambda^{(1)}\ad\tau_-^2 -\Lambda^{(1)}\ad\tau_+^2\right)+\mrm{H.c.}\Big],\notag\\
	H_\mrm{SB}^2(t)\simeq&-f^{(2)}(t)\Big[\lambda^{(2)}\ad\tau_-^2+\Lambda^{(2)}\ad\tau_+^2\\
		&-\hat\lambda_\mrm J\left(\lambda^{(2)}\ad\tau_-^1+\Lambda^{(2)}\ad\tau_+^1\right)\tau_z^2+\mrm{H.c.}\Big],\notag
\end{align}
corresponding to red and blue sidebands. In practice, these terms will average out unless the modulation frequency $\omega_\mrm{FC}$ is chosen appropriately. For example, taking $\omega_\mrm{FC}^{(k)}\simeq\Delta^{(k)}$ or $\omega_\mrm{FC}^{(k)}\simeq\Sigma^{(k)}$, respectively lead to red or blue sideband transitions between the resonator and qubit $k$. These transitions respectively correspond to $\ket{1;n}\leftrightarrow\ket{0;n+1}$ and $\ket{1;n+1}\leftrightarrow\ket{0;n}$, with the first number labeling the state of qubit $k$ and the second being a Fock state of the resonator. Additionally, because of the qubit-qubit interaction $J$, we also get leakage-like terms which are proportional to $\hat\lambda_\mrm J$: driving qubit $k$ at the frequency corresponding to a sideband for qubit $k'$ will drive the sideband in qubit $k'$. In this process, qubit $k$ will pick up a phase. While these terms could be useful for logical gates, they are in practice very small because of the $\hat\lambda_\mrm J$ factor which is of second order in $g^{(k)}$. Furthermore, the assumption of a large qubit-qubit detuning makes these leakage terms far off-resonant with the other sideband terms, making these spurious transitions negligible in practice. We also point out that in most cases, $\lambda^{(k)} \gg \Lambda^{(k)}$, implying that the red sidebands rate is much larger than for the blue sideband.

As in the single qubit case, we also unsurprisingly find a term which corresponds to parametric amplification~\cite{PhysRevA.84.043832}
\be
	H_\mrm{PO}(t)\simeq-\sum_{k=1,2}\lambda^{(k)}\Lambda^{(k)}f^{(k)}(t)\;\tau_z^k(a^2+\ad\,\!^2).
\ee
This term is only relevant for $\omega_\mrm{FC}\simeq2\omega_r$. However, in typical situations, it is not large enough to be useful as a practical source of squeezed microwave light~\footnote{The standard deviation in the $X$ quadrature of a degenerate parametric oscillator when $\omega_\mrm{FC}\simeq2\omega_r$ is $\Delta X=\frac14\eul{-2r}$, where $r=\frac12\mrm{arctanh}(4\epsilon^{(k)}\lambda^{(k)}\Lambda^{(k)}/\omega_r)$~\cite{walls-milburn}. For the generalized dispersive transform to be valid, we require $\lambda^{(k)},\Lambda^{(k)}\ll1$. Therefore, $r\gtrsim1$ implies $\epsilon^{(k)}\gg\omega_r$, which is impossible if $\omega_a\sim\omega_r$ as is usually the case in circuit QED.}.

Finally, the virtual qubit-qubit interaction is responsible for the appearance of a new term in the FC Hamiltonian in the diagonal frame
\begin{align}
	&H_\mrm{QQ}(t) \simeq \Big\{-\frac{f^{(1)}(t)+f^{(2)}(t)}{2}\left(\lambda^{(1)}\lambda^{(2)}-\Lambda^{(1)}\Lambda^{(2)}\right)\notag\\
				&\hspace{1.25cm}- \hat\lambda_\mrm J \left[f^{(1)}(t)-f^{(2)}(t)\right]\Big\}\left(\tau_-^1\tau_+^2+\tau_+^1\tau_-^2\right)\\
				& -\frac{f^{(1)}(t)+f^{(2)}(t)}{2}\left(\lambda^{(1)}\Lambda^{(2)}-\Lambda^{(1)}\lambda^{(2)}\right)\left(\tau_-^1\tau_-^2+\tau_+^1\tau_+^2\right).\notag
\end{align}
Because of this term, when $\omega_\mrm{FC}\simeq\Delta_\mrm Q$, a flip-flop interaction between the two logical qubits is turned on (first two lines of $H_\mrm{QQ}$). This can be used to generate $\mrm{\sqrt{SWAP}}$ gates to entangle the qubits. Special care must however be taken when it comes to the relative phase between the modulation functions. Indeed, taking $f^{(1)}(t)$ and $f^{(2)}(t)$ in phase results in a cancelation of the term proportional to $f^{(1)}(t)-f^{(2)}(t)$, while taking them to be out of phase cancels the term in $f^{(1)}(t)+f^{(2)}(t)$. Therefore, depending on the qubit and resonator frequencies, one should choose the phase that allows to select the fastest population transfer. We also point out that if $\omega_a^{(1)}<\omega_r<\omega_a^{(2)}$, the signs of $\lambda^{(1)}\lambda^{(2)}$ and $\Lambda^{(1)}\Lambda^{(2)}$ are opposite, allowing a larger Rabi frequency if $\Lambda^{(k)}$'s are comparable to $\lambda^{(k)}$'s. We can also drive a transition that creates or destroy a quantum in each qubit when $\omega_\mrm{FC}\simeq\Sigma_\mrm Q$ (last term line of $H_\mrm{QQ}$). Because of its $g^4$ dependance, the effect of $H_\mrm{QQ}$ is small for the parameters chosen in this paper and will be ignored in the following.

\section{Sideband control with FC drives \label{sec:SB}}

Since they are the terms of largest amplitudes in $H_\mrm{FC}^\mrm{diag}$ that can lead to two-qubit gates, we  focus here on the red sideband terms present in the first lines of $H^1_\mrm{SB}$ and $H^2_\mrm{SB}$. As already pointed out, with respect to previous proposals~\cite{blais2007quantum} and realizations~\cite{PhysRevB.79.180511,PhysRevLett.104.100504} in circuit QED, we emphasize that the sideband rate under an FC drive is of first order in $g$, rather than of second order. In this section, we calculate the evolution of the state of a target qubit $k$ and of the resonator under a red sideband transition generated by an FC drive on $k$, the other qubit $k'$ merely acting as a spectator.  However, because of the qubit-induced frequency pull of the resonator, the resonance frequency for a sideband on the target qubit $k$ depends (weakly) on the state of the spectator qubit $k'$. This leads to a small error that increases with $\lambda^{k'}_{i,i+1}$, which is evaluated below. Finally, because our goal is to see how good a qubit-qubit entangling gate based on the red sideband could be in principle, we will ignore here the effect of damping and dephasing. These will simply lower the fidelities in a straightforward manner.

As discussed in Section~\ref{sec:tunable}, at the appropriate FC frequency, red sideband transitions occur between eigenstates in the diagonal dispersive frame. These eigenstates can be written in the basis $\left\{\ket{\tau_1\,\tau_2;n}\right\}$, where $\tau_k\in\{0,1\}$ denotes the eigenstates of logical qubit $k$ and $n$ represents Fock states of the dispersively-shifted resonator.  Focussing on red sideband transitions, which preserve the total number of quanta, the Hilbert space can be restricted to the subspace $\{{\ket{00;n+1}},{\ket{10;n}},{\ket{01;n+1}},{\ket{11;n}}\}$. To simplify the discussion, we keep only the relevant terms in the total transformed Hamiltonian in the presence of an FC drive at the red sideband frequency. This leads to
\begin{align}
	&H_\mrm{red}(t)=\omega_r\ad a+\sum_k \frac{\tilde\omega_{a}^{(k)}+2\ad a S^{(k)}}{2}\tau_z^{(k)}\notag\\
		&\qquad-\lambda^{(1)}f^{(1)}(t)\left(\ad\tau_-^{(1)}+a\tau_+^{(1)}\right).\label{eq:Hred}
\end{align}
This is a good approximation when $\lambda^{(1)}\varepsilon^{(1)}\ll |\tilde\omega_a^{(1)}-\omega_r|,\;|\tilde\omega_a^{(2)}-\omega_r|$, and $|\tilde\omega_a^{(1)}-\tilde\omega_a^{(2)}|$, i.e. the transitions that can be stimulated by the FC drive are well separated. Furthermore, we have neglected $H_z^1(t)$ and $H_z^2(t)$ which correspond to oscillations in the qubit frequencies on a timescale that is much smaller than the sideband transitions. These terms can be safely dropped if the average of $f^{(1)}(t)$ during a whole period is zero, which is the case here with a drive described by Eq.~\eq{eq:fk}.

It is useful to move to the interaction picture with respect to 
\be
	U_\mrm{rot}(t) = \exp\left[-i\left(\omega_r\ad a + \sum_k\frac{\tilde\omega_{a}^{(k)}+2\ad a S^{(k)}}{2}\tau_z^{(k)}\right)t\right].
\ee
The resulting Hamiltonian is
\begin{align}
	&H_\mrm{red}'(t)=-\lambda^{(1)}f^{(1)}(t) \left[a\tau_+^{(1)}\exp\left(i\hat\Delta^{(1)}_nt\right)+\mrm{H.c.}\right],
\end{align}
where $\hat\Delta^{(1)}_{n\pm}=\tilde\omega_a^{(1)}-\omega_r+2S^{(1)} \ad a - S^{(2)} \tau_z^{(2)}$. Working in the restricted Hilbert space and ignoring dissipation, Schr\"odinger's equation for the above Hamiltonian and with the drive defined by Eq.~\eq{eq:fk}, leads to two uncoupled sets of differential equations for the probably amplitudes $c^\pm_{0/1}(t)$ and corresponding to the second qubit being excited ($+$) or not ($-$)
\begin{align}
	\dot c^\pm_0(t) &=ic^\pm_1(t)\epsilon_n\left[\eul{i\left(\omega_\mrm{FC}-\Delta^\pm_n\right)t}+\eul{-i\left(\omega_\mrm{FC}+\Delta^\pm_n\right)t}\right]\\
	\dot c^\pm_1(t) &=ic^\pm_0(t)\epsilon_n\left[\eul{-i\left(\omega_\mrm{FC}-\Delta^\pm_n\right)t}+\eul{i\left(\omega_\mrm{FC}+\Delta^\pm_n\right)t}\right],
\end{align}
where $\omega_\mrm{FC}\equiv\omega_\mrm{FC}^{(1)}$ and 
\be
	\Delta^\pm_n = \left(\tilde\omega_a^{(1)}+2(n+1)S^{(1)}\right)-\left(\omega_r\pm S^{(2)}\right),
\ee
\be
	\epsilon_n = \frac12\lambda^{(1)}\varepsilon^{(1)}\sqrt{n+1}\label{eq:eps:n}.
\ee
For $\epsilon_n\ll\omega+\Delta_n^\pm$, the fast oscillating terms can be neglected and an analytic solution can be found. The evolution operator corresponding to this solution is, for each subspace $\pm$
\be
\begin{split}
&	V^\pm(t) 
	\\
&	=\matdeux{\eul{-i\delta_\pm t/2}\left[\cos \frac{rt}{2}-i\frac{\delta_\pm}{r}\sin \frac{rt}{2}\right]&-2i\frac{\epsilon_n}{r}\eul{-i\delta_\pm t/2}\sin \frac{rt}{2}\\ -2i\frac{\epsilon_n}{r}\eul{i\delta_\pm t/2}\sin \frac{rt}{2} & \eul{i\delta_\pm t/2}\left[\cos \frac{rt}{2}+i\frac{\delta_\pm}{r}\sin \frac{rt}{2}\right]}\label{eq:V},
\end{split}
\ee
where $r=\sqrt{\delta_\pm^2+4\epsilon^2_n}$ and $\delta_\pm=\omega_\mrm{FC}-\Delta^\pm_n$ is the detuning between the FC drive and the Stark-shifted resonance frequency of the red sideband. We already notice that without coupling between the second qubit and the resonator, we eliminate the Stark shift and thus can have $\delta_+=\delta_-=0$. This leads to full population inversion between ${\ket{0;n+1}}$ and ${\ket{1;n}}$ at the Rabi frequency $2\epsilon_n$, which is of first order in $g^{(k)}/\Delta^{(k)}$.

Given the evolution operator $V^\pm(t)$, we now calculate the average fidelity to a red sideband $\pi$-pulse on the first qubit, irrespective of the state of the second qubit. The ideal process is described by
\be
	U=-i\left(\ket{0;n+1}\bra{1;n}+\ket{1;n}\bra{0;n+1}\right)\otimes\id_{Q2}.
\ee
To compute the average fidelity, it is useful to define the Choi matrices corresponding respectively to the ideal $C_\mrm U$ and the actual $C_\mrm V(t)$ processes~\cite{choi1975completely,kraus1983states}
\begin{align}
	C_\mrm U&=\frac12\sum_{i,j\in\left\{a,b\right\}}\ket i\bra j_c\otimes U\ket i\bra j U^\dagger\label{eqn:choi:U}\\
	C_\mrm V(t)&=\frac12\sum_{i,j\in\left\{a,b\right\}}\ket i\bra j_c\otimes V(t)\ket i\bra j V^\dagger(t).\label{eqn:choi:V}
\end{align}
 The sums above are evaluated over states $\ket{a}=\ket{0;1}$ and $\ket{b}=\ket{1;0}$ that are involved in the red sideband quantum process. The subsystem labelled $c$ in the above equations represents a copy of the relevant Hilbert space $\{\ket{0;1},\ket{1;0}\}$ and is required in the definition of the Choi matrix.
The gate fidelity $F_\mrm{UV}(t)=\mathrm{tr}\left[C_\mrm U C_\mrm V(t)\right]$ of that process is thus 
\be\label{eqn:trace}
	F_\mrm{UV}(t)=\frac14\left|\mrm{tr}\left[U^\dagger V(t)\right]\right|^2.
\ee
Using Eq.~\eq{eqn:choi:U} and~\eq{eqn:choi:V}, we obtain
\be\label{eqn:FUV}
	F_\mrm{UV}(t)=2\frac{\epsilon_n^2}{r^2}\sin^2\frac{rt}{2}\left(1+\cos\delta_\pm t\right),
\ee
from which we obtain the average fidelity~\cite{MichaelA2002249}
\be
	\overline F_\mrm{UV}=\frac{2F_\mrm{UV}+1}{3}.\label{eqn:avg:fid}
\ee
As expected, the fidelity is unity with $\epsilon_n t=\pi/2$ and if the Stark shift $S^{(2)}$ due to the second qubit vanishes. In practice, $S^{(2)}$ reduces the average fidelity.

\begin{figure}
	\begin{center}
		\includegraphics[scale=0.7]{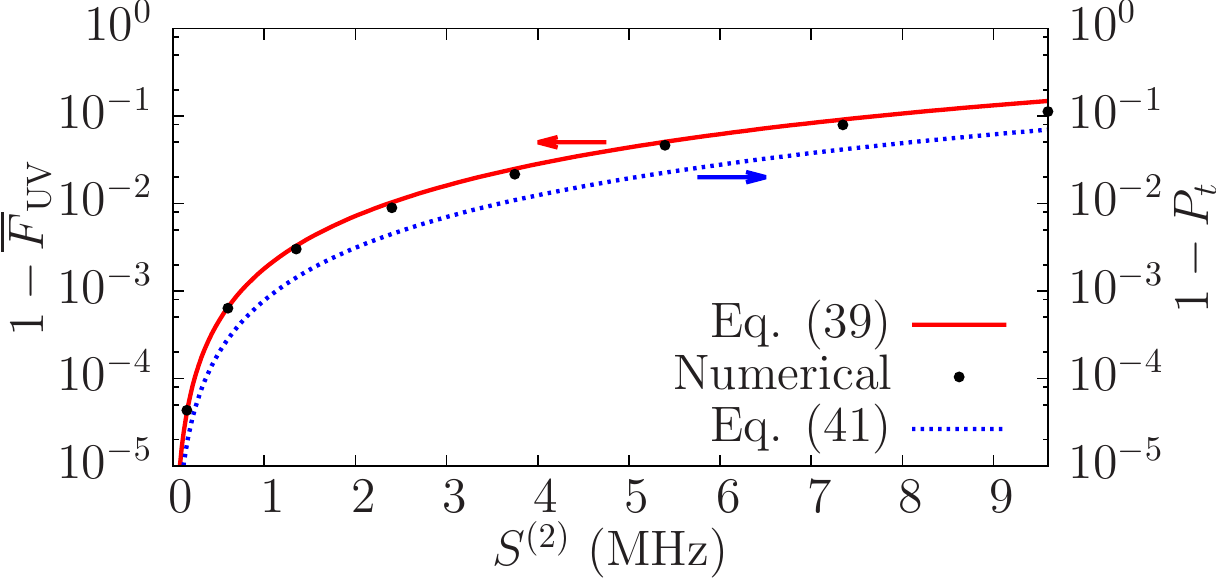}
	\end{center}
	\caption{(Color online) Average error with respect to the perfect red sideband process $\ket{1;0}\leftrightarrow\ket{0;1}$. A gaussian FC pulse is sent on the first qubit at the red sideband frequency assuming the second qubit is in its ground state. Full red line: average error of the red sideband as given by Eq.~\eq{eqn:FUV:simple} when the second qubit is excited. Blue dashed line: population transfer error $1-P_t$, with $P_t$ given by Eq.~\eq{eqn:pop:transfer}. Black dots: numerical results for the average error. We find the evolution operator after time $t_p$ for each eigenstate of the second qubit. The fidelity is extracted by injecting these unitaries in Eq.~\eq{eqn:trace}. The qubits are taken to be transmons, which are modelled as 4-level Duffing oscillators (see Section~\ref{sec:Duffing}) with $E_J^{(1)}=25$ GHz, $E_J^{(2)}=35$ GHz, $E_C^{(1)}=250$ MHz, $E_C^{(2)}=300$ MHz, yielding $\omega_{01}^{(1)}=5.670$ GHz and $\omega_{01}^{(2)}=7.379$ GHz, and $g_{01}^{(1)}=100$ MHz. The resonator is modeled as a 5-level truncated harmonic oscillator with frequency $\omega_r=7.8$ GHz. As explained in Section~\ref{sec:transmon}, the splitting between the first two levels of a transmon is modulated using a time-varying external flux $\phi$. Here, we use gaussian pulses in that flux, as described by Eq.~\eq{eqn:gaussian} with $\tau=2\sigma$, $\sigma=6.6873$ns, and flux drive amplitude $\Delta\phi=0.075\,\phi_0$. The length of the pulse is chosen to maximize the population transfer.\label{fig:FUV}}
\end{figure}

As an example, we now assume the modulation frequency to be at the red sideband transition frequency, given that  the second qubit is in its ground state. In other words, we have
\be
	\delta_\pm = \left\{\begin{array}{l} 0\mbox{ if the second qubit is in state }\ket{0}\\ 2S^{(2)}\mbox{ if the second qubit is in state }\ket{1} \end{array}\right..
\ee
In this situation and for a given temporal shape of the FC modulation, the condition for population inversion becomes $\overline \epsilon_n t=\pi/2$, with
\be
	\overline\epsilon_n=\frac{1}{t_p}\int_0^{t_p}\epsilon_n(t)dt.
\ee
Taking $\overline \epsilon_n t_p=\pi/2$ and replacing $\epsilon_n\rightarrow\overline\epsilon_n$ in Eq.~\eq{eqn:FUV} yields a simple expression for the gate fidelity in the limit $\delta_+\ll\overline\epsilon_n$
\be
	F_\mrm{UV}\simeq\frac{2\overline\epsilon^2_n}{\delta_+^2+4\overline\epsilon^2_n}\left(1+\cos\delta_+t_p\right).\label{eqn:FUV:simple}
\ee
In words, the infidelity $1-F_\mrm{UV}$ is minimized when the Rabi frequency that corresponds to the FC drive is large compared to the Stark shift associated to the spectator qubit. The average fidelity corresponding to the gate fidelity Eq.~\eq{eqn:FUV:simple} is illustrated in Fig.~\ref{fig:FUV} as as a function of $S^{(2)}$ (red line) assuming the second qubit to be in its excited state. 
We also represent as black dots a numerical estimate of the error coming from the spectator qubit's Stark shift. The latter is calculated with Eqs.~\eq{eqn:trace} and~\eq{eqn:avg:fid}. Numerically solving the system's Schr\"odinger equation allows us to extract the unitary evolution operator that corresponds to the applied gate. Taking $U$ to be that evolution operator for the spectator qubit in state $\ket 0$ and $V$ the operator in state $\ket1$, we obtain the error caused by the Stark shift shown in Fig.~\ref{fig:FUV}. The numerical results closely follow the analytical predictions, even for relatively large dispersive shifts $S^{(2)}$.

Reducing the dispersive shift of the spectator qubit $S^{(2)}$ rapidly increases the average fidelity. There are several ways to do this. The straightforward solution is to bring it to a very large detuning with the resonator, thus reducing the Stark shift. One could also use a tunable coupling~\cite{PhysRevLett.105.023601,PhysRevLett.106.030502}. Finally, if neither of these options is convenient, this kind of issue can typically be addressed by pulse shaping techniques~\cite{PhysRevA.77.032315}.

It should also be noted that if one is interested only in population transfers and not in  phase information, the above analysis exaggerates the error because it cares about phase. In this context, a more appropriate figure of merit is the population of the target state
\be
	P_t=\left|\bra{1;n}V(t)\ket{0;n+1}\right|^2=\frac{4\overline{\epsilon}_n^2}{r^2}\sin^2\left(\frac{rt}{2}\right).
\ee
Taking, as in the previous case, $t=t_p=\pi/2\overline\epsilon_n$ and expanding to second order in $\delta_+/\overline\epsilon_n$, this reduces to
\be\label{eqn:pop:transfer}
	P_t\simeq\frac{4\overline{\epsilon}_n^2}{\delta_+^2+4\overline\epsilon_n^2}.
\ee
This expression is the blue dotted line in Fig.~\ref{fig:FUV}. Clearly, not taking phases into account yields a much smaller error rate. As will be discussed in Section~\ref{sec:CNOT}, this feature remarkably helps the implementation of two-qubit gates,  Indeed, in that case, the specific phase of each sideband gate is irrelevant, provided that it is well-controlled and stable over many repetitions of the experiment.

\section{Many-level systems and logical gates based on sideband control \label{sec:mls}}

While the two-level model presented in the previous sections is useful because of its simplicity, it is not an appropriate description of most superconducting qubits, such as the transmon~\cite{koch2007charge} or the CSFQ~\cite{PhysRevB.75.140515}. Indeed, these devices are best described as many-level systems. In this section, we generalize the previous results to that more realistic situation and discuss how these additional levels can by exploited to our advantage. As above, we start be obtaining an effective Hamiltonian using a dispersive transformation and then diagonalize the remaining qubit-qubit interaction. We only outline the main steps of this calculation, the details of which can be found in Appendix~\ref{sec:disp:mls}, before presenting a sequence of pulses corresponding to a qubit-qubit entangling gate that takes advantage of the many-level structure of the qubits.

\subsection{FC drive Hamiltonian for many-level systems\label{sec:FC:MLS}}

The Hamiltonian of two many-level systems (MLS) coupled to the same harmonic mode takes the form~\cite{koch2007charge}
\begin{align}
	&H=\omega_r\ad a + \sum_{k=1,2}\sum_{i\in\mathcal H_M}\omega_i^{(k)}\Pi_{i,i}^{(k)} \label{eqn:H:MLS}\\
		&\qquad+ \sum_{k=1,2}\sum_{i\in\mathcal H_{M-1}}g_i^{(k)}\left(\Pi_{i,i+1}^{(k)}\ad+\Pi_{i,i+1}^{(k)}a\right)+ \mrm{H.c.}.\notag
\end{align}
with $\Pi_{i,j}^{(k)}=\ket i\bra j^{(k)}$ acting on the $k$-th MLS, $\omega_i^{(k)}$ the bare MLS frequencies,  $g_i^{(k)}$ the coupling rate between states $i$ and $i+1$ of the $k$-th MLS and the resonator, and $M$ the number of levels in each MLS. $\mathcal H_M=\{\ket g,\ket e,\ket f,\ket h,...\}$ is the Hilbert space of a bare $M$-level system. For simplicity, we have assumed that only nearest-neighbor states are coupled by the electric-dipole interaction. As will be discussed further below, this approximation is well satisfied for transmons~\cite{koch2007charge}.

As shown in Appendix~\ref{sec:disp:mls}, the dispersive transformation described by Eq.~\eqref{eqn:tls:transform} can be generalized to the multilevel case. This transformation $U_\mrm D^\mrm{MLS}$, given by Eq.~\eq{eq:transform}, is used to approximately diagonalize the above Hamiltonian, leading to Stark shifts. As in the simple case, this also leads to a qubit-qubit coupling term of amplitude $J$. An additional transformation $U_J^\mrm{MLS}$ is introduced to diagonalize this effective coupling. Successively applying $U_\mrm D^\mrm{MLS}$ and $U_J^\mrm{MLS}$, we find to second order in $g_i^{(k)}$
\begin{align}
	&H_\mrm D^\mrm{diag} =\omega_r\ad a+\sum_{k=1,2}\sum_{i=0}^{M-1}\left[\tilde{\omega}_i^{(k)}\Pi_{i,i}^{(k)}+S_i^{(k)}\Pi_{i,i}^{(k)}\ad a\right],\label{eqn:diag:MLS}
\end{align}
where 
\begin{align}
	\tilde\omega_i^{(k)}&=\omega_i^{(k)}+L_i^{(k)},\\
	L_i^{(k)}&=\chi_{i-1}^{(k)}-\mu_{i}^{(k)},\\
	S_i^{(k)}&=\chi_{i-1}^{(k)}+\mu_{i-1}^{(k)}-\chi_i^{(k)}-\mu_i^{(k)}.
\end{align}
Here, we have defined $\chi_i^{(k)}=g_i^{(k)\ast}\lambda_i^{(k)}$ and $\mu_i^{(k)}=g_i^{(k)\ast}\Lambda_i^{(k)}$, respectively the dispersive and Bloch-Siegert shifts with $\lambda_i^{(k)}=g_i^{(k)}/\Delta_i^{(k)}$, $\Lambda_i^{(k)}=g_i^{(k)}/\Sigma_i^{(k)}$, $\Delta_i^{(k)}=\omega_{i+1}^{(k)}-\omega_i^{(k)}-\omega_r$, and $\Sigma_i^{(k)}=\omega_{i+1}^{(k)}-\omega_i^{(k)}+\omega_r$. The eigenstates of the above Hamiltonian are, in each photon subspace, logical (dressed) transmon states $\ket{\pi_1\pi_2;n}$, where $\pi_k\in\{0,1,2,...\}$.

We now introduce the FC drive on each MLS. The most general Hamiltonian that describes this in the bare frame is
\be
	H_\mrm{FC}=\sum_{k=1,2}\sum_{i\in\mathcal H_M}f_i^{(k)}(t)\Pi_{i,i}^{(k)},
\ee
where $f_i^{(k)}(t)$ are real functions of time. Successively applying $U_\mrm D^\mrm{MLS}$ and $U_J^\mrm{MLS}$ on that Hamiltonian, we obtain an expression for $H_\mrm{FC}$ in the frame that diagonalizes Eq.~\eq{eqn:H:MLS}. To simplify the discussion, we drop here terms that are of second order in $g_i^{(k)}$ because they lead to tractable but unsightly expressions. In this way, we find that the first-order corrections are sideband terms
\begin{align}
	H_\mrm{FC}^\mrm{diag}\simeq H_\mrm{FC}-\sum_{k=1,2}\sum_{i=0}^{M-2}H_\mrm{SB,i}^{(k)}, \label{eqn:frst:order}
\end{align}
where
\be
	H_\mrm{SB,i}^{(k)}=\delta f_i^{(k)}(t)\left[\lambda_i^{(k)}\Pi_{i,i+1}^{(k)}\ad+\Lambda_i^{(k)}\Pi_{i,i+1}^{(k)}a+\mrm{H.c.}\right],
\ee
and $\delta f_i^{(k)}(t)=f_{i+1}^{(k)}(t)-f_i^{(k)}(t)$.

\begin{figure}
	\begin{center}
		\includegraphics[scale=0.8]{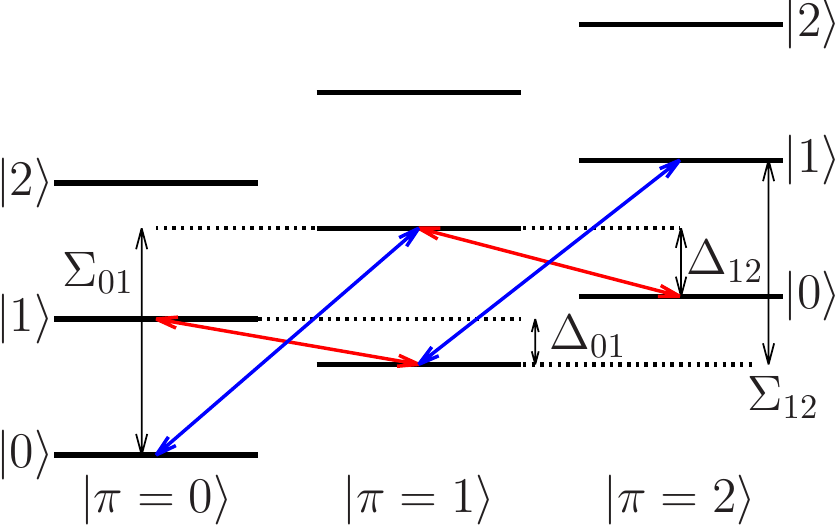}
	\end{center}
	\caption{(Color online) Sideband transitions for a three-level system coupled to a resonator. Applying an FC drive at frequency $\Delta_{i,i+1}$ generates a red sideband transitions between states $\ket{i+1;n}$ and $\ket{i;n+1}$, where the numbers represent respectively the MLS and resonator states. Similarly, driving at frequency $\Sigma_{i,i+1}$ leads to a blue sideband transition, i.e. $\ket{i;n}\leftrightarrow\ket{\mbox{$i+1;n+1$}}$. Transitions between states higher in the Fock space are not shown for reasons of readability. This picture is easily generalized to an arbitrary number of levels.\label{fig:MLS:sidebands}}
\end{figure}

Thus, instead of having only one red and one blue sideband transitions for each photon number as in the two-level case,  here $M-1$ red and blue sidebands are  possible for each $n$. These are  illustrated in Fig.~\ref{fig:MLS:sidebands}. For large enough anharmonicity of the MLSs, these sidebands are at distinct frequencies. As will be discussed in the next section, this selectivity is a useful resource to design quantum logical gates.

\subsection{Pulse sequence generating a maximally entangling gate\label{sec:pulse:sequence}}

As already mentioned, sideband transitions can be used to generate entanglement between two qubits coupled to the same resonator. Here, we show how the second excited dressed level of a MLS, labelled $\ket 2$, can be exploited to design a maximally entangling gate.

Let us define $\sigma_{x,i,j}^{(k)}=\Pi_{i,j}^{(k)}+\Pi_{j,i}^{(k)}$, corresponding to a $\pi$-pulse between the levels $i$ and $j$ of the $k$-th logical MLS, and $R_{i,i+1}^{(k)}=\Pi_{i,i+1}^{(k)}\ad+\Pi_{i+1,i}^{(k)}a$, corresponding to a full red sideband transition between the $k$-th MLS and the resonator. Using these definitions, we introduce the pulse sequence
\be
	U_\mrm{ent}=R_{01}^{(1)}R_{12}^{(2)}\sigma_{x,12}^{(2)}R_{12}^{(2)}R_{01}^{(1)}.\label{eq:pulse}
\ee
Up to phases, this is equivalent to the CNOT. Indeed, looking at the effects of this sequence on each state of the two-qubit computational basis $\left\{\ket{00},\ket{10},\ket{01},\ket{11}\right\}$ given that the resonator is initially in its ground state and ignoring phases for the moment, we have
\begin{align*}
	&\ket{00;0} \rightarrow \ket{00;0} \rightarrow \ket{00;0} \rightarrow \ket{00;0} \rightarrow \ket{00;0} \rightarrow \ket{00;0}\\
	&\ket{10;0} \rightarrow \ket{00;1} \rightarrow \ket{00;1} \rightarrow \ket{00;1} \rightarrow \ket{00;1} \rightarrow \ket{10;0}\\
	&\ket{01;0} \rightarrow \ket{01;0} \rightarrow \ket{01;0} \rightarrow \ket{02;0} \rightarrow \ket{01;1} \rightarrow \ket{11;0}\\
	&\ket{11;0} \rightarrow \ket{01;1} \rightarrow \ket{02;0} \rightarrow \ket{01;0} \rightarrow \ket{01;0} \rightarrow \ket{01;0}.
\end{align*}
Therefore, the first qubit is flipped if the second one is in state $\ket 1$. This is represented more generally in matrix form by
\be
	U_\mrm{ent}=\matquatre{1 & 0 & 0 & 0\\ 0 & \eul{i\phi_{1}} & 0 & 0\\ 0 & 0 & 0 & \eul{i\phi_{2}}\\ 0 & 0 & \eul{i\phi_3} & 0},\label{eqn:U:ent}
\ee
where $\phi_1$, $\phi_2$, and $\phi_3$ are phases that depend on the shape of the FC pulses. Indeed, as will be seen in Section~\ref{sec:FC:transmon}, in realistic situations, the FC drive is not strictly a cosine drive on all the qubit transitions frequencies, but also has a component that can be interpreted as a constant shift on these frequencies, leading to additional phase accumulations.

The above unitary is equivalent, up to one-qubit gates, to the CNOT. Indeed, we can write
\be
	\mrm{CNOT}=U^{(1)}_{\theta_1}U_\mrm{ent}U^{(2)}_{\theta_2}U^{(1)}_{\theta_3},
\ee
where $U^{(k)}_{\theta}$ is a rotation of qubit $k$ along the $z$-axis by an angle $\theta$ and
\begin{align}
	\theta_1&=(\phi_2-\phi_1-\phi_3)/2,\\
	\theta_2&=(\phi_1-\phi_2-\phi_3)/2,\\
	\theta_3&=(\phi_3-\phi_1-\phi_2)/2.
\end{align}

It should be noted that the above sequence should be faster than the one presented in Ref.~\cite{blais2007quantum}, since it contains four sideband pulses instead of five. Together with the fact that sidebands generated with FC drives have a first-order nature, this allows much faster results than expected in the initial proposal.

\section{Physical implementation with transmons}
\label{sec:transmon}

In the previous sections, we first introduced FC drives on two-level systems, and then of generic MLSs. Here, we focus on a specific kind of MLS: the transmon qubit~\cite{koch2007charge,schreier2008suppressing}. Using a simple model, we predict that the red sideband transition can be generated with a very high Rabi frequency using an FC drive on that qubit. We also show simulation results for the pulse sequence presented in Section~\ref{sec:pulse:sequence}.

\subsection{The transmon as a Duffing oscillator\label{sec:Duffing}}

A simple Hamiltonian for the transmon qubit can be derived from the Cooper-pair box Hamiltonian by expanding it to fourth order in $E_C/E_J$. In the limit where $E_C/E_J\ll1$, the system is well approximated by a Duffing oscillator~\cite{koch2007charge}
\be	\label{eqn:duffing}
	H_\mrm{trans}\simeq \sqrt{8E_C E_J}\bd b-\frac{E_C}{12}\left(b+\bd\right)^4,
\ee
where $E_J$ and $E_C$ are respectively the Josephson and capacitive energy of the transmon, while $b$ and $\bd$ are the ladder operators of this weakly anharmonic oscillator. The associated eigenfrequencies are then
\be
	\omega_j \simeq \left(\omega_p-\frac{E_C}{2}\right)j-\frac{E_C}{2}j^2.	\label{eqn:freqs}
\ee
with $\omega_p=\sqrt{8E_CE_J}$ the plasma frequency which depends on $E_J$. 
Additionally, the transmon-resonator coupling rates are, again in the regime of large $E_J/E_C$ ratios, given by
\begin{align}
	g_{j,j+1}&\simeq g_{ge}\sqrt{j+1},\\
	g_{j,k} &\simeq 0 \quad\forall\quad k\neq j\pm1,
\end{align}
where
\be
	g_{ge}=-2i\beta eV^0_\mrm{rms}\frac{1}{\sqrt2}\left(\frac{E_J}{8E_C}\right)^{1/4}.
\ee
In addition, $g_{i+1,i}=g_{i,i+1}^\ast$. The transmon-resonator couplings also depend weakly on $E_J$.

\subsection{FC drives on the transmon-resonator system\label{sec:FC:transmon}}

If the transmon's Josephson junction is replaced by a SQUID, $E_J$ can be tuned with an external flux. Assuming the flux modulation to have the form of a cosine wave, we have
\be	\label{eqn:flux:drive}
	E_J(t)=E_{J\Sigma}\cos\left[\phi_\mrm{i}+\Delta\phi\cos(\omega_\mrm{FC}t)\right],
\ee
where $\phi_i$ and $\Delta\phi$ are the mean value and amplitude of the external flux drive. All the fluxes are expressed in units of $ \Phi_0/\pi$, where $\Phi_0$ is the flux quantum. This results in modulations of the plasma frequency and thus in the transmon energies, something which can be used for FC driving. Additionally, this also leads to modulation of the transmon-resonator couplings. These modulations qualitatively have the same effect as the transmon frequency modulation, leading to red and blue sidebands. However, because of the weak dependence of the coupling on $E_J$, this is a small effect which can safely be dropped.

\begin{figure}
	\begin{center}
		\vspace{-2mm}
		\includegraphics[scale=0.65]{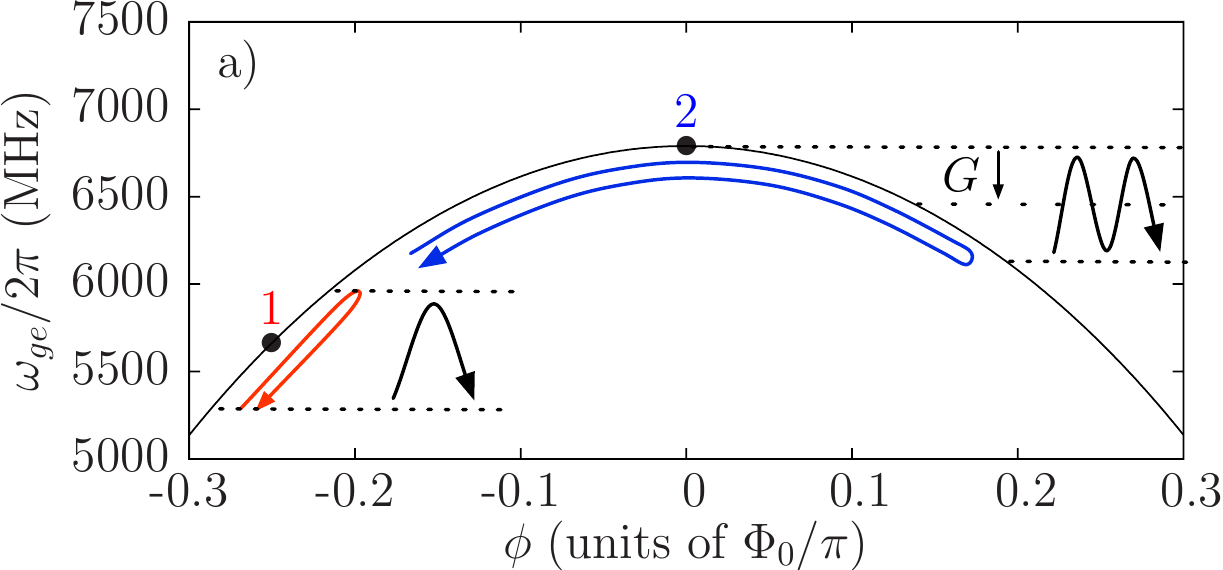}\\
		\includegraphics[scale=0.65]{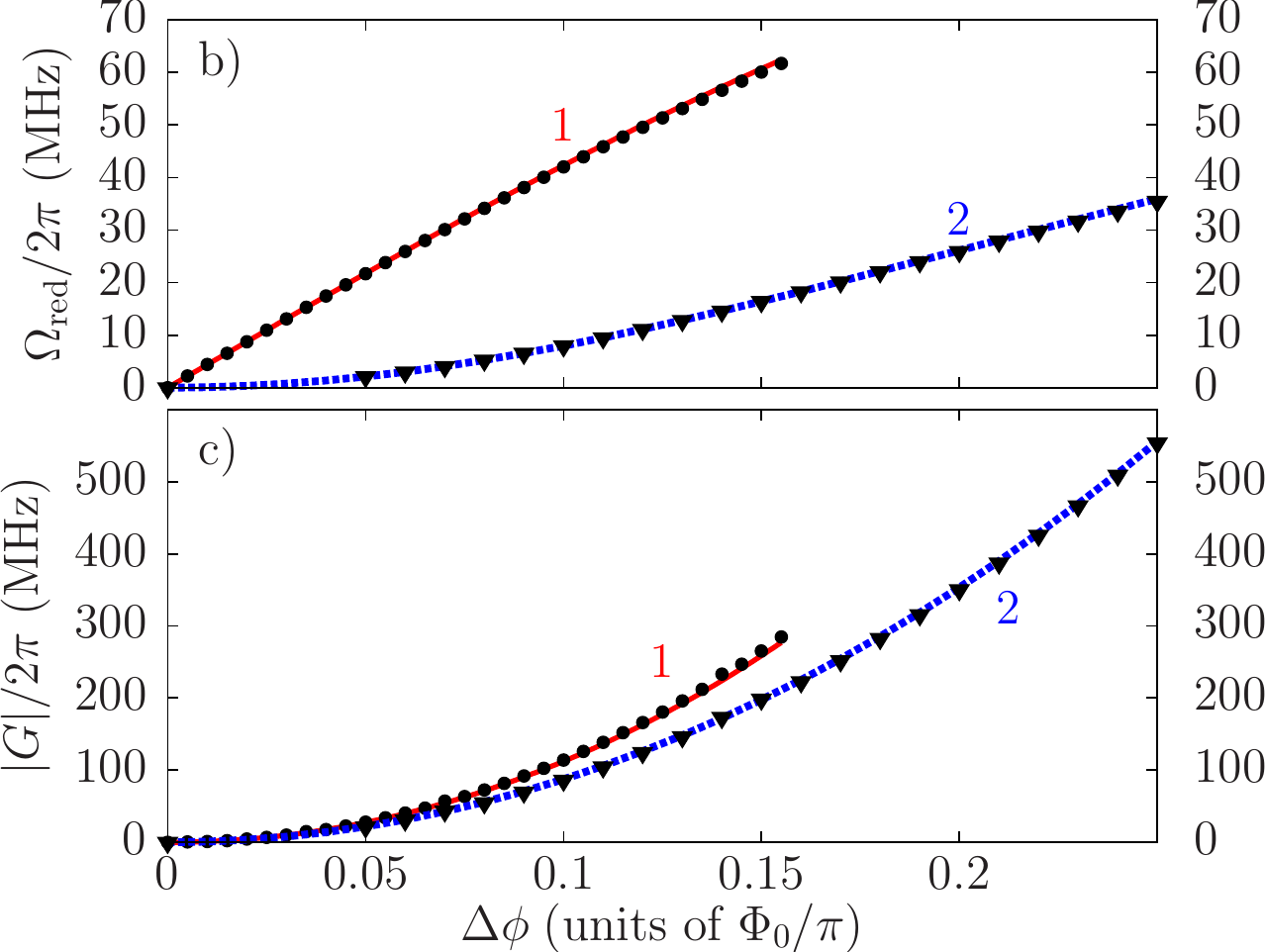}\\
		\includegraphics[scale=0.6]{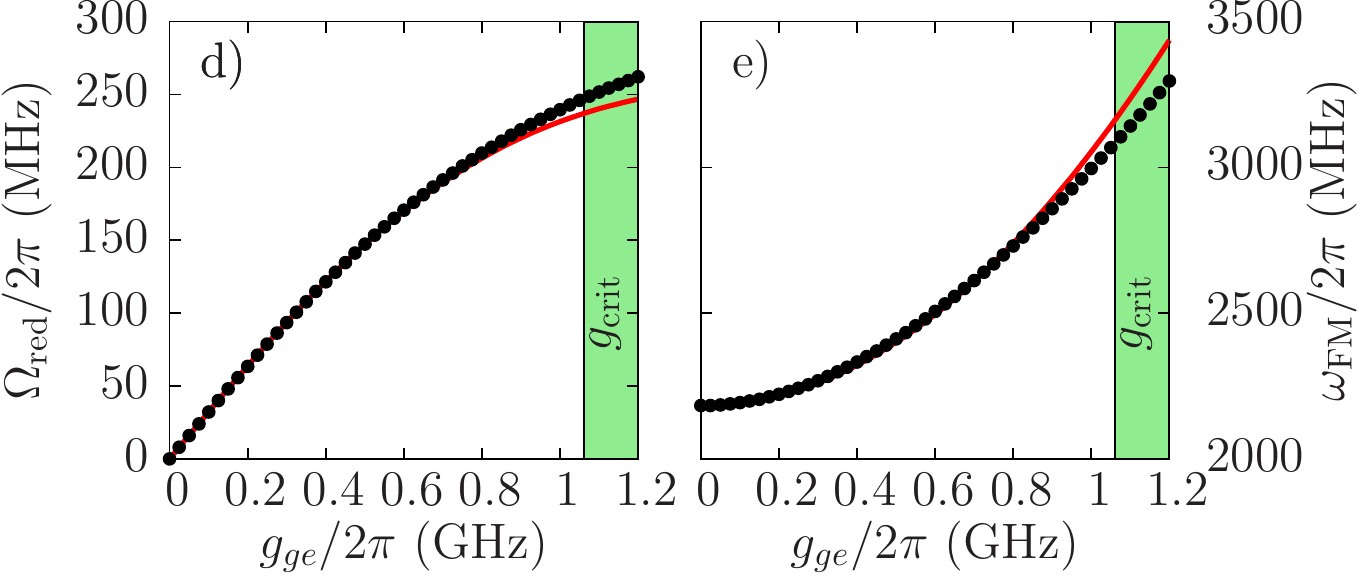}
		\vspace{-3mm}
	\end{center}
	\caption{(Color online) FC driving of a transmon with an external flux. The transmon is modelled using the first four levels of the Hamiltonian given by Eq.~\eq{eqn:duffing}, using parameters $E_J/2\pi=25$ GHz and $E_C/2\pi=250$ MHz. We also have $g_{ge}/2\pi=100$ MHz and $\omega_r/2\pi=7.8$ GHz, which translates to $\Delta_{ge}/2\pi\simeq2.1$ GHz. a) Frequency of the transition to the first excited state obtained by numerical diagonalization of Eq.~\eq{eqn:duffing}. As obtained from Eqs.~\eq{eqn:hamonic:1} to~\eq{eqn:hamonic:4}, the major component in the spectrum of $\omega_{ge}(t)$ when shaking the flux away from the flux sweet spot at frequency $\omega_\mrm{FC}$ also has frequency $\omega_\mrm{FC}$. However, when shaking around the sweet spot, the dominant harmonic has frequency $2\omega_\mrm{FC}$. Furthermore, the mean value of $\omega_{ge}$ is shifted by $G$. b) Rabi frequency of the red sideband transition $\ket{1;0}\leftrightarrow\ket{0;1}$. The system is initially in $\ket{1;0}$ and evolves under the Hamiltonian given by Eq.~\eq{eqn:H:MLS} and a flux drive described by Eq.~\eq{eqn:flux:drive}. Full red line: analytical results from Eq.~\eq{eqn:rabi:freq} with $m=1$ and $\phi_i=0.25$. Dotted blue line: $m=2$ and $\phi_i=0$. Black dots and triangles: exact numerical results. c) Geometric shift for $\phi_i=0.25$ (full red line) and 0 (dotted blue line). d) Increase in the Rabi frequency for higher coupling strengths with $\phi_i=0.25$ and $\Delta\phi=0.075$. e) Behavior of the resonance frequency for the flux drive. As long as the dispersive approximation holds ($g\lesssim g_\mrm{crit}/2\pi=1061$ MHz), it remains well approximated by Eq.~\eq{eqn:resonance}, as shown by the full red line. The same conclusion holds for the Rabi frequency. \label{fig:transmon}}
\end{figure}

Having dropped the modulations of the coupling, we now focus on better understanding the frequency modulations. Expanding Eq.~\eq{eqn:flux:drive} around $\phi_i$ to fourth order in $\Delta\phi$ and for $\phi_i$ such that $\tan\phi_i\lesssim1$, i.e. $\phi_i\lesssim\pi/4$,  we obtain the following formula for the transmon transition frequencies 
\begin{align}\label{eq:TransmonFreqModulation}
	\omega'_{j,j+1}(t) \simeq \omega_{j,j+1}+G-\sum_{m=1}^4\varepsilon_{m\omega}\cos m\omega_\mrm{FC}t,
\end{align}
where $\omega_{j,j+1} = \omega_{j+1} - \omega_{j}$ is the transition frequency without the external modulation
\be
	\omega_{j,j+1}=\omega_p'-E_C(j+1).
\ee
We have also defined $\omega'_p=\sqrt{8E_CE_{J\Sigma}\cos\phi_i}$, the plasma frequency associated to the operating point $\phi_i$. This frequency is illustrated by the black dots for two operating points on Fig.~\ref{fig:transmon}a). In addition, there is a frequency shift $G$,  standing for \emph{geometric}, that depends on the shape of the transmon energy bands. As is also illustrated on Fig.~\ref{fig:transmon}a), this frequency shift  comes from the fact that the relation between $\omega_{j,j+1}$ and $\phi$ is nonlinear, such that the mean value of the transmon frequency during flux modulation is not its value for the mean flux $\phi_i$. To fourth order in $\Delta\phi$, it is
\begin{align}
	G\simeq&-\left(1+\frac{\tan^2\phi_i}{2}\right)\omega_p'\frac{\Delta\phi^2}{8}\notag\\
		&-\left(4+20\tan^2\phi_i+15\tan^4\phi_i\right)\omega_p'\frac{\Delta\phi^4}{1024}.\label{eq:G}
\end{align}
Finally, we get in Eq.~\eqref{eq:TransmonFreqModulation} a modulation at each multiple $m\omega_\mrm{FC}$ of the flux drive frequency. The amplitude of these modulations diminishes with increasing $m$ as
\begin{align}
	\varepsilon_\omega\simeq&\left[\Delta\phi+\left(1+\frac32\tan^2\phi_i\right)\frac{\Delta\phi^3}{16}\right]\omega'_p\frac{\tan\phi_i}{2},\label{eqn:hamonic:1}\\
	\varepsilon_{2\omega}\simeq&\left(1+\frac{\tan^2\phi_i}{2}\right)\omega_p'\frac{\Delta\phi^2}{8}\notag\\
		&+\left(4+20\tan^2\phi_i+15\tan^4\phi_i\right)\omega_p'\frac{\Delta\phi^4}{768},\label{eqn:hamonic:2}\\
	\varepsilon_{3\omega}\simeq&\left(\frac13+\frac{\tan^2}{2}\right)\omega_p'\tan\phi_i\frac{\Delta\phi^3}{32},\label{eqn:hamonic:3}\\
	\varepsilon_{4\omega}\simeq&\left(4+20\tan^2\phi_i+15\tan^4\phi_i\right)\omega_p'\frac{\Delta\phi^4}{3072}\label{eqn:hamonic:4}.
\end{align}
As illustrated in Fig.~\ref{fig:transmon}a), in the special case where $\phi_i=0$, i.e. around the flux sweet spot, only the even harmonics $2\omega_d$ and $4\omega_d$ are non-zero. Furthermore, in realistic situations, $\Delta \phi \ll 0.1$, meaning that $\epsilon_{m\omega}$ falls off extremely quickly with $m$. This allows to focus only on the first non-vanishing harmonic coming from the FC drive in the following discussion.

Let $m$ be that dominant harmonic. Then, shaking the flux at a frequency such that 
\be \label{eqn:resonance}
	m\omega_\mrm{FC}=\tilde\Delta_{j,j+1}^n+G,
\ee
it is possible to induce red sideband transitions between states $\ket{\mbox{$j+1;n$}}\leftrightarrow\ket{j;n+1}$ at the Rabi frequency
\begin{align}	\label{eqn:rabi:freq}
	\Omega_{j,j+1}\simeq\left|\frac{g_{j,j+1}}{\tilde\Delta_{j,j+1}^n+G}\right|\varepsilon_{m\omega},
\end{align}
where $\tilde\Delta_{j,j+1}^n=\tilde\omega_{j+1}-\tilde\omega_j-\omega_r+n(S_{j+1}-S_j)-S_j$ is the Lamb and Stark-shifted detuning between the resonator frequency and the $j$-th transmon splitting.

In Fig.~\ref{fig:transmon}b), the Rabi frequencies predicted by the above formula are compared to numerical simulations using the full Hamiltonian Eq.~\eq{eqn:H:MLS}, along with a cosine flux drive. The geometric shifts described by Eq.~\eq{eq:G} are also plotted in Fig.~\ref{fig:transmon}c), along with numerical results. In both cases, the scaling with respect to $\Delta\phi$ follows very well the numerical predictions, allowing us to conclude that our simple analytical model accurately synthesizes the physics occurring in the full Hamiltonian. It should be noted that, contrary to intuition, the geometric shift is roughly the same at and away from the sweet spot. This is simply due to the fact that the band curvature does not change much between the two operation points. However, as expected from Eqs.~\eq{eqn:hamonic:1} to~\eq{eqn:hamonic:4}, the Rabi frequencies are much larger for the same drive amplitude when the transmon is on average away from its flux sweet spot. In that regime, large Rabi frequencies $\sim 30$-40 MHz can be attained, which is well above dephasing rates in actual circuit QED systems, especially in the 3D cavity~\cite{PhysRevLett.107.240501}. However, the available power that can be sent to the flux line might be limited in the lab, putting an upper bound on achievable rates. Furthermore, at those rates, fast rotating terms such as the ones dropped between Eq.~\eq{eq:eps:n} and~\eq{eq:V} start to play a role, adding spurious oscillations in the Rabi oscillations that reduce the fidelity. These additional oscillations have been seen to be especially large for big relevant $\varepsilon_{m\omega}/\tilde\Delta_{j,j+1}^n$ ratios, i.e. when the qubit spends a significant amount of time close to resonance with the resonator and the dispersive approximation breaks down.

Moreover, the analytical model developed here relies on the dispersive approximation, and thus breaks down unless $\lambda_{j,j+1}^{(k)}$ is small. This breakdown is illustrated in Fig.~\ref{fig:transmon}d) and \ref{fig:transmon}e), which respectively illustrate the behavior of the Rabi frequency and the geometric shift with increasing $g^{(k)}$. The regime of validity of the dispersive approximation can be captured semi-quantitatively using the critical photon number $n_\mrm{crit}=\Delta^2/4g^2$~\cite{blais2004cavity}. Inverting that relation, we get a critical coupling strength $g_\mrm{crit}=\Delta/2\sqrt{n}$. For a red sideband involving only one excitation, this reduces to $g_\mrm{crit}=\Delta/2$. The region for which $g>g_\mrm{crit}$ corresponds to the shaded green areas in the figure. It correctly corresponds to the point where our analytical model starts to deviate from the the full numerical results.

\subsection{CNOT gate using FC driving on transmons \label{sec:CNOT}}

In this Section, we present simulation results for the pulse sequence described in Eq.~\eq{eq:pulse} using the red sideband described above for the transmon. The envelope function of the FC flux modulation is chosen to be a truncated Gaussian 
\be\label{eqn:gaussian}
	\Delta\phi(t) = \left\{\begin{array}{l}A \eul{-(t-\mu)^2/2\sigma^2}\mbox{ if } \mu-\tau\leq t\leq\mu+\tau,\\
							     0\mbox{ otherwise.} \end{array}\right.
\ee
In order to reduce leakage to unwanted levels due to the weak anharmonicity of the transmon, we use first-order DRAG corrections on $R_{12}^{(2)}$ and $\sigma_{x,12}^{(2)}$ pulses~\cite{PhysRevA.83.012308}. More precisely, for each corrected Gaussian pulse, we add a drive component that is phase-shifted by $\pi/2$ and that is proportional to the derivative of Eq.~\eq{eqn:gaussian} for $|t-\mu|<\tau$. 

\begin{figure}
	\begin{center}
		\includegraphics[scale=0.75]{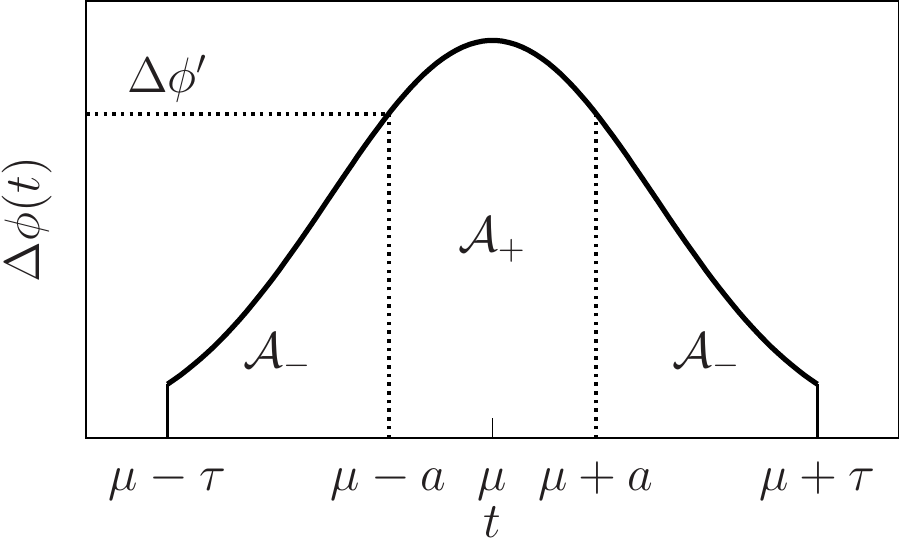}
	\end{center}
	\caption{Amplitude of the gaussian pulse over time. $\Delta\phi'$ is such that the areas $\mathcal A_+$ and $2\mathcal A_-$ are equal. Then, driving the sideband at its resonance frequency for the geometric shift that corresponds to the flux drive amplitude $\Delta\phi'$ allows population inversion. \label{fig:gaussian}}
\end{figure}

In practice, using a device such as a transmon, we do not only have leakage out of the computational subspace, but we also have to take into account the geometric shift described in the preceding section. Indeed, since that shift depends on the flux drive amplitude, the resonance frequency of the red sideband process changes during the application of FC drives with $G(\Delta\phi(t))$. In order to cancel that effect, we drive the sideband at the resonance frequency for a geometric shift that corresponds to $\Delta\phi'$, as illustrated on Fig.~\ref{fig:gaussian}. This amplitude is chosen such that the pulse area for which the geometric shift is above $G(\Delta\phi')$ is the same as the pulse area for which it is below $G(\Delta\phi')$. Using this definition yields $\Delta\phi'=A\exp(-a^2/2\sigma^2)$ where
\be
	a=\sqrt2\sigma\;\mrm{erf}^{-1}\left[\frac12\mrm{erf}\left(\frac{\tau}{\sqrt2\sigma}\right)\right].
\ee
\begin{table}
	\begin{center}
		\begin{ruledtabular}
			\begin{tabular}{ccccc}
				Pulse & $A[\phi_0/\pi,2\pi\times\mrm{MHz}]$ & $\mu$[ns] & $\sigma$[ns] & $2\tau$[ns]\\
				\hline
				$R_{01}^{(1)}$ & 0.07308 & 16 & 7 & 28\\
				$R_{12}^{(2)}$ & 0.02520 & 46.5 & 6.25 & 25\\
				$\sigma_{x,12}^{(2)}$ & 51.1412 & 65.96 & 1.48 & 5.92\\
				$R_{12}^{(2)}$ & 0.02520 & 85.42 &6.25 & 25\\
				$R_{01}^{(1)}$ & 0.07308 & 115.92 & 7 & 28				
			\end{tabular}
		\end{ruledtabular}
	\end{center}
	\caption{Pulse sequence used in the simulations of a CNOT using transmons. The amplitude of the pulses are in units of $\phi_0/\pi$ for the flux pulses and in MHz for the direct drive on the second transmon. The total duration of the sequence is 129.92ns. Transmon parameters are $E_J^{(1)}/2\pi=25$GHz, $E_C^{(1)}/2\pi=250$MHz, $\phi^{(1)}=0.25$, $E_J^{(2)}/2\pi=61$GHz, $E_C^{(2)}/2\pi=300$MHz, $\phi^{(2)}=0.25$, $g_{ge}^{(1)}/2\pi=g_{ge}^{(2)}/2\pi=100$MHz. Four levels are considered for each transmon. The resonator is truncated to 5 levels and has $\omega_r/2\pi=7.8$GHz.
\label{tab:sequence}}
\end{table}

This method is first applied to simulate a $R_{01}^{(1)}$ pulse by evolving the two-transmon-one-resonator system under the Hamiltonian of Eq.~\eq{eqn:H:MLS}, along with the FC drive Hamiltonian for the pulse. The simulation parameters are indicated in Table~\ref{tab:sequence}. To generate the sideband pulse $R_{01}^{(1)}$, the target qubit splitting is modulated at a frequency that lies exactly between the red sideband resonance for the spectator qubit in states $\ket{0}$ or $\ket{1}$, such that the fidelity will be the same for both these spectator qubit states. We calculate the population transfer probability for $\ket{1;0}\leftrightarrow\ket{0;1}$ after the pulse and find a success rate of 99.2\% for both initial states $\ket{1;0}$ and $\ket{0;1}$. This is similar to the prediction from Eq.~\eq{eqn:pop:transfer}, which yields 98.7\%. The agreement between the full numerics and the simple analytical results is remarkable, especially given that with $|\delta_\pm/\epsilon_n|=0.23$ the small $\delta_\pm\ll\epsilon_n$ assumption is not satisfied. Thus, population transfers between the transmon and the resonator are achievable with a good fidelity even in the presence of Stark shift errors coming from the spectator qubit (see Section~\ref{sec:SB}).

The pulse sequence $U_\mrm{ent}$ of Eq.~\eqref{eq:pulse} is then used to numerically prepare qubit-qubit entangled states. The parameters of every pulses entering in $U_\mrm{ent}$ are presented in Table~\ref{tab:sequence}. Given that as many as four sideband pulses are used, one might expect the success probability to be rather low. However, applying that sequence on each of the separable states $(\ket{00}\pm\ket{01})/\sqrt{2}$, $(\ket{10}\pm\ket{11})/\sqrt{2}$, we optain overlaps with maximally entangled states of $\sim99\%$, as shown in Table~\ref{tab:bell}. The phases coming from the geometric shifts, which appear in Eq.~\eq{eqn:U:ent} describing $U_\mrm{ent}$, are found by choosing the set $\{\phi_1,\phi_2,\phi_3\}$ that optimizes the transfer rates. 
\begin{table}
	\begin{center}
		\begin{ruledtabular}
			\begin{tabular}{ccccc}
				Init. state & $P(\ket{\Phi_+})$ & $P(\ket{\Phi_-})$ & $P(\ket{\Psi_+})$ & $P(\ket{\Psi_-})$\\
				\hline
				$\ket{\phi_+}$	& $2\times10^{-4}$ & \textbf{0.993} & $2\times10^{-5}$ & $8\times10^{-5}$\\
				$\ket{\phi_-}$	& $\mathbf{0.993}$ & $8\times10^{-5}$ & $6\times10^{-5}$ & $8\times10^{-5}$\\
				$\ket{\psi_+}$	& $8\times10^{-5}$ & $9\times10^{-5}$ & $7\times10^{-5}$ & $\mathbf{0.991}$\\
				$\ket{\psi_-}$	& $5\times10^{-5}$ & $3\times10^{-5}$ & $\mathbf{0.978}$ & $3\times10^{-5}$\\	
			\end{tabular}
		\end{ruledtabular}
	\end{center}
	\caption{Population transfer succes rates from the basis of product states to the basis of Bell states using the pulse sequence displayed in Table~\ref{tab:sequence}. The phases are $\phi_1=0.053,\,\phi_2=2.31$, and $\phi_3=5.59$. 
\label{tab:bell}}
\end{table}

The average fidelity is found numerically by using as initial state the Choi matrix $\sum_{jk}\ket j \bra k_c \otimes\ket j \bra k$, with $j,k\;\in\;\{\ket{00;0},\ket{01;0},\ket{10;0},\ket{11;0}\}$. The evolution is computed numerically on the first copy under the Hamiltonian of Eq.~\eq{eqn:H:MLS}, along with the drive Hamiltonians for the pulses. In this way, we find the set of phases $\{\phi_1,\phi_2,\phi_3\}$ that best fits the numerically realized gate and find an average fidelity of $99.1\%$. It should be noted that the presence of decoherence can only degrade these fidelities.  In order to obtain better fidelities one would need to use more sophisticated pulse shaping techniques to compensate for the Stark shift errors.

\begin{figure}
	\begin{center}
		\includegraphics[scale=0.69]{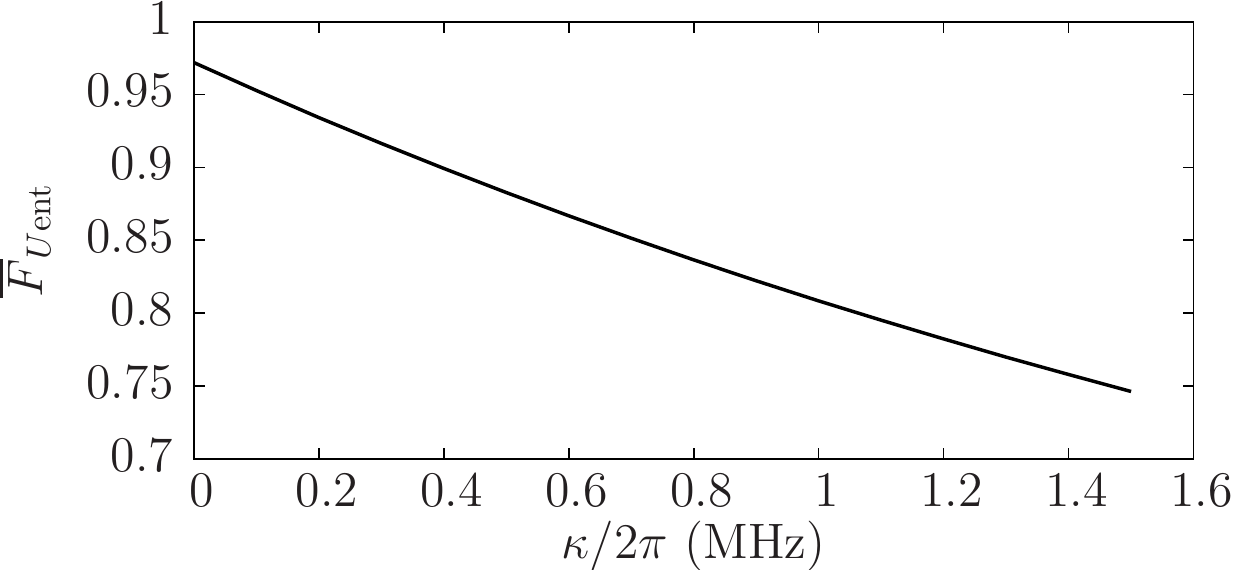}
	\end{center}
	\caption{Average fidelity of the pulse sequence $U_\mrm{ent}$ as a function of cavity damping damping $\kappa$. The simulation includes realistic damping $T_1=2\mu$s of the transmons and neglects pure dephasing. Purcell decay is taking into account in the simulation. The other parameters are given in Table~\ref{tab:sequence}. \label{fig:FvsK}}
\end{figure}

Finally, we included dissipation in our model using a Markovian master equation for two MLS and a resonator~\footnote{We use Eq. (2.6) of Ref.~\cite{PhysRevA.85.022305} with $\kappa_\mrm{NL}=\gamma_\phi=0$.}. Because the resonator is loaded with real (and not only virtual) photons, we find that the protocol is sensitive to photon loss $\kappa$. Fig.~\ref{fig:FvsK} shows the average fidelity as a function of the cavity decay rate. High gate fidelities require low cavity damping. This is possible using 3D cavities~\cite{PhysRevLett.107.240501} or taking advantage of the multi-mode structure of 2D resonators~\cite{PhysRevLett.104.100504}. Note that the dependence of the fidelity on qubit relaxation and dephasing is similar.

\section{Discussion}

\begingroup
\squeezetable
\begin{table*}
	\begin{center}
		\begin{ruledtabular}
		\begin{tabular}{lcccccc}
			  Scheme & Operation & Theoretical rate & Crossings & On/off ratio\\
			  \hline
			  Capacitive coupling~\cite{bialczak2010quantum} & $\sqrt{i\mrm{SWAP}}$ & $\sim J_\mrm C$ & Yes & $\sim\Delta_Q^2/J_\mrm C^2$ \\
			  11-02 anti-crossing~\cite{dicarlo2009demonstration,dicarlo2010preparation} & CPHASE & $\sim J_{11-02}$ & Yes & $\sim\left|\omega_{ef}^{(k)}-\omega_{ge}^{(k')}\right|^2/J_{11-02}^2$ \\
			  Cross-resonance~\cite{PhysRevLett.107.080502} & CNOT & $\sim J\epsilon/\Delta_Q$ & None & $\infty$ \\
			  2$^\mrm{nd}$-order blue SB~\cite{PhysRevB.79.180511,PhysRevLett.104.100504} & Bell state gen. & $\sim g^3\epsilon^2/\Delta^4$ & None & $\infty$ \\
			  1$^\mrm{st}$-order red SB & CNOT & $\sim g\epsilon/\Delta$ & None$^\ast$ & $\infty$ \\
		\end{tabular}
		\end{ruledtabular}
		\caption{Schemes for two-qubit operations in circuit QED. $\epsilon$ is the strength of the drive used in the scheme, if any. $(^\ast)$ There are no crossings in that gate provided that the qubits have frequencies separated enough that they do not overlap during FC modulations. \label{tab:gates}}
	\end{center}
\end{table*}
\endgroup

Table~\ref{tab:gates} summarizes theoretical predictions and experimental results for recent proposals for two-qubit gates in circuit QED. These can be divided in two broad classes. The first includes approaches that rely on anticrossings in the qubit-resonator or qubit-qubit spectrum. They are typically very fast, since their rate is equal to the coupling strength involved in the anticrossing. Couplings can be achieved either through direct capacitive coupling of the qubits with strength $J_\mrm{C}$~\cite{bialczak2010quantum}, or through the 11-02 anticrossing in the two-transmon spectrum which is mediated by the cavity~\cite{dicarlo2009demonstration, dicarlo2010preparation}. The latter technique has been successfully used with large coupling rates $J_{11-02}$ and  Bell-state fidelities of $\sim94\%$. However, since these gates are activated by tuning the qubits in and out of resonance, they have a finite on/off ratio determined by the distance between the relevant spectral lines. Thus, the fact that the gate is never completely turned off will make it very complicated to scale up to large numbers of qubits. Furthermore, adding qubits in the resonator leads to more spectral lines that also reduce scalability. In that situation, turning the gates on and off by tuning qubit transition frequencies in and out of resonance without crossing these additional lines becomes increasingly difficult as qubits are added in the resonator, an effect known as spectral crowding. 

The second class of gates includes cross-resonance, second-order sidebands, and our proposal. Since these approaches rely on external drives, their rate is not strictly set by the relevant coupling rate between involved states, but also by the ratio between the drive strength and some detuning, which in practice is always smaller than 1. This makes these gates slightly slower than those relying on anticrossings.
Since these approaches rely on external drives that can be turned on and off at will, their on/off ratio is in principle infinite. Therefore, even if they take longer times than anticrossing approaches, they will become more interesting as $T_2$ gets better. Additionally, since they are operated either with fixed qubit frequencies or with qubit splittings oscillating around a central frequency, they should be less subject to spectral crowding. It should be noted that our proposal is predicted to be faster than cross-resonance, since it scales with $g$ rather than with $J$, which is of second order in $g$.

\begin{acknowledgments}
	Special thanks to Thomas A. Ohki, Blake R. Johnson, Colm A. Ryan, Maxime Boissonneault, J\'er\^ome Bourassa, and William A. Coish for fruitful discussion. F.B. was funded by NSERC and FQRNT. A.B. acknowledges funding from NSERC, the Alfred P. Sloan Foundation, and CIFAR. F.B., M.P.S., and Z.D. acknowledge support from IARPA under contract W911NF-10-1-0324. All statements of fact, opinion or conclusions contained herein are those of the authors and should not be construed as representing the official views or policies of the U.S. Government.
\end{acknowledgments}

\appendix

\section{Diagonalization of the two-qubit-one-resonator system in the dispersive regime\label{sec:diag}}

As shown in Sec.~\ref{sec:gen:disp}, after the application of the generalized dispersive transformation of Eq.~\eqref{eqn:tls:transform}, the second-order two-qubit plus one resonator Hamiltonian Eq.~\eqref{eqn:J:Hamiltonian} is not completely diagonal. In this appendix, we diagonalize the remaining non-diagonal term which corresponds to virtual qubit-qubit interaction. 

This is done by first realizing that the Hamiltonian of Eq.~\eqref{eqn:J:Hamiltonian} breaks down into $4\times 4$ blocks corresponding to a given Fock state $n$. One such block is given in Eq.~\eq{eqn:blocks} and can be diagonalized exactly. Doing so, we get four eigenenergies and four eigenstates per photon subspace. These eigenstates span a basis labeled $\{\ket{00;n},\ket{01;n},\ket{10;n},\ket{11;n}\}$ that defines two effective uncoupled qubits whose frequencies depend on the resonator state. Expressed as a function of the eigenstates in the dispersive basis, these effective qubit states are
\begin{subequations}
\begin{align}
	\ket{10;n}&=-\sin\alpha_n\ket{gen}+\cos\alpha_n\ket{egn},\label{eq:diag:1}\\
	\ket{01;n}&=\cos\alpha_n\ket{gen}+\sin\alpha_n\ket{egn},\\
	\ket{00;n}&=\cos\beta_n\ket{ggn}+\sin\beta_n\ket{een},\\
	\ket{11;n}&=-\sin\beta_n\ket{ggn}+\cos\beta_n\ket{een}\label{eq:diag:4},
\end{align}
\end{subequations}
with the mixing angles
\begin{subequations}
\begin{align}
	\alpha_n&=\arctan\left[\frac{\tilde\Delta_\mrm Q^n/2-\sqrt{J^2+(\tilde\Delta_\mrm Q^n/2)^2}}{J}\right],\label{eq:mix:1}\\
	\beta_n&=\arctan\left[\frac{\tilde\Sigma_\mrm Q^n/2-\sqrt{J^2+(\tilde\Sigma_\mrm Q^n/2)^2}}{J}\right]\label{eq:mix:2}.
\end{align}
\end{subequations}
The corresponding energies are
\begin{align}
	E_{10/01}&=\pm\sqrt{J^2+(\tilde\Delta_\mrm Q^n/2)^2}\\
	E_{11/00}&=\pm\sqrt{J^2+(\tilde\Sigma_\mrm Q^n/2)^2}.
\end{align}
Defining the Pauli operators $\tau_z^1=(\proj1-\proj0)\otimes\mathbb{I}^{(2)}$ and $\tau_z^2=\mathbb{I}^{(1)}\otimes(\proj1-\proj0)$, we can express the diagonalized two-qubit plus one-resonator Hamiltonian as
\begin{widetext}
\begin{equation}	
\begin{split}
H_\mrm{diag}  = \omega_r\ad a & + \left(\sqrt{J^2+(\Sigma_\mrm Q/2+\ad a\Sigma_\mrm S)^2}+\sqrt{J^2+(\Delta_\mrm Q/2+\ad a\Delta_\mrm S)^2}\right)\frac{\tau_z^1}{2}\\
			&+\left(\sqrt{J^2+(\Sigma_\mrm Q/2+\ad a\Sigma_\mrm S)^2}-\sqrt{J^2+(\Delta_\mrm Q/2+\ad a\Delta_\mrm S)^2}\right)\frac{\tau_z^2}{2}.
\end{split}
\end{equation}
\end{widetext}
In the regime where $J\ll\Sigma_\mrm Q+2n\Sigma_\mrm S,\Delta_\mrm Q+2n\Delta_\mrm S$, i.e. at large qubit-qubit detuning and low photon number, we can expand the square roots to second order in $J/(\Sigma_\mrm Q+2n\Sigma_\mrm S)$ and $J/(\Delta_\mrm Q+2n\Delta_\mrm S)$ to obtain the approximate Hamiltonian of Eq.~\eq{eq:Hdiag:disp}.

Using Eq.~\eq{eq:diag:1} to~\eq{eq:diag:4}, it is useful to express the Pauli matrices in the dispersive frame $\sigma^j_{\pm,z}$ in terms of the effective qubit Pauli matrices $\tau^j_{\pm,z}$
\begin{align}
	\sz^1=&\frac{\cos2\hat\alpha+\cos2\hat\beta}{2}\tau_z^1-\frac{\cos2\hat\alpha-\cos2\hat\beta}{2}\tau_z^2\notag\\
		&+2\sin\hat\alpha\cos\hat\alpha\left(\tau_-^1\tau_+^2+\tau_+^1\tau_-^2\right)\notag\\
		&+2\sin\hat\beta\cos\hat\beta\left(\tau_-^1\tau_-^2+\tau_+^1\tau_+^2\right),\label{eq:sz1}\\
	\sz^2=&-\frac{\cos2\hat\alpha-\cos2\hat\beta}{2}\tau_z^1+\frac{\cos2\hat\alpha+\cos2\hat\beta}{2}\tau_z^2\notag\\
		&-2\sin\hat\alpha\cos\hat\alpha\left(\tau_-^1\tau_+^2+\tau_+^1\tau_-^2\right)\notag\\
		&+2\sin\hat\beta\cos\hat\beta\left(\tau_-^1\tau_-^2+\tau_+^1\tau_+^2\right)\label{eq:sz2},
\end{align}
\begin{align}
	\sigma_\pm^1=&\cos\hat\alpha\cos\hat\beta\,\tau_\pm^1-\sin\hat\alpha\sin\hat\beta\,\tau_\mp^1\notag\\
		&-\cos\hat\alpha\sin\hat\beta\,\tau_z^1\tau_\mp^2-\sin\hat\alpha\cos\hat\beta\,\tau_z^1\tau_\pm^2,\label{eqn:disp2diag:spm}\\
	\sigma_\pm^2=&\cos\hat\alpha\cos\hat\beta\,\tau_\pm^2+\sin\hat\alpha\sin\hat\beta\,\tau_\mp^2\notag\\
		&-\cos\hat\alpha\sin\hat\beta\,\tau_\mp^1\tau_z^2+\sin\hat\alpha\cos\hat\beta\,\tau_\pm^1\tau_z^2,
\end{align}
where $\hat\alpha$ and $\hat\beta$ are operators obtained from the mixing angles $\alpha_n$ and $\beta_n$ by replacing $n$ by $\ad a$. 

Using these expressions, we can now express any control drive on the qubits in the diagonal frame. As a first example, it is useful to consider a direct drive on qubit~1 which is described in the bare frame by
\be
	H_{d1}=\varepsilon_{d1}(t)\left(\sm^1\eul{i\omega_{d1}}+\mrm{H.c.}\right).
\ee
Ignoring dispersive corrections, in the frame rotating at the qubit and resonator frequencies and assuming $\omega_{d1}=\omega_a^{(2)}$, we find an additional term of the form
\be
	H_\mrm{cross}=-\varepsilon_{d1}(t)\left(\cos\hat\alpha\sin\hat\beta+\sin\hat\alpha\cos\hat\beta\right)\tau_z^1\tau_x^2.
\ee
This corresponds to the cross-resonance gate~\cite{PhysRevB.81.134507,PhysRevLett.107.080502}. The present  description has the advantage of including the effects of the resonator state. Indeed, we find that the Rabi frequency associated to this gate depends on the photon number. For $J\ll\Delta_\mrm Q$ and neglecting counter-rotating terms, this rate is proportional to $J/(\Delta_\mrm Q+2n\Delta_\mrm S)$. Therefore, depending on the sign of $\Delta_\mrm S$ with respect to $\Delta_\mrm Q$, additional photons act as an increased or decreased qubit-resonator detuning. If $\Delta_\mrm S$ and $\Delta_\mrm Q$ have the same sign, which happens for $\omega^{(1)}_a,\omega^{(2)}_a<\omega_r$ or $\omega^{(1)}_a,\omega^{(2)}_a>\omega_r$, the cross resonance rate increases when the cavity is filled with photons. Otherwise, that rate decreases when photons are added, allowing us to expect a saturation of the cross-resonance rate as a function of the drive power. 

We now express the FC drives in the diagonalized basis. Expressing the Hamiltonian of Eq.~\eq{eqn:fm} in the diagonal frame using Eq.~\eq{eq:sz1} and~\eq{eq:sz2}, we find
\begin{align}
	H_\mrm{FC}^\mrm{diag}(t)=&\;H_z^1(t)+H_z^2(t)+H_\mrm{SB}^1(t)+H_\mrm{SB}^2(t)\notag\\
		&+H_\mrm{PO}(t)+H_\mrm{QQ}(t)+H^\phi_\mrm{QQ}(t)\label{eq:FC:exact:H},
\end{align}
where the effects of the $J$ coupling are taken into account exactly. In this expression, we have defined
\begin{widetext}
\begin{align}
	H_z^1(t)&=\left\{\hat s_n^{(1)}f^{(1)}(t)\left[\cos2\hat\alpha+\cos2\hat\beta\right]-\hat s_n^{(2)}f^{(2)}(t)\left[\cos2\hat\alpha-\cos2\hat\beta\right]\right\}\frac{\tau_z^1}{4},\\
	H_z^2(t)&=\left\{-\hat s_n^{(1)}f^{(1)}(t)\left[\cos2\hat\alpha-\cos2\hat\beta\right]+\hat s_n^{(2)}f^{(2)}(t)\left[\cos2\hat\alpha+\cos2\hat\beta\right]\right\}\frac{\tau_z^2}{4},
\end{align}
\begin{align}
	H_\mrm{SB}^1(t)=&-f^{(1)}(t)\left[\left(\lambda^{(1)}\cos\hat\alpha\cos\hat\beta-\Lambda^{(1)}\sin\hat\alpha\sin\hat\beta\right)\ad\tau_-^1 + \left(\Lambda^{(1)}\cos\hat\alpha\cos\hat\beta-\lambda^{(1)}\sin\hat\alpha\sin\hat\beta\right)\ad\tau_+^1 \right.\notag\\
	&-\left.\left(\lambda^{(1)}\cos\hat\alpha\sin\hat\beta+\Lambda^{(1)}\sin\hat\alpha\cos\hat\beta\right)\ad\tau_z^1\tau_+^2-\left(\lambda^{(1)}\sin\hat\alpha\cos\hat\beta+\Lambda^{(1)}\cos\hat\alpha\sin\hat\beta\right)\ad\tau_z^1\tau_-^2+\mrm{H.c.}\right],\\
	H_\mrm{SB}^2(t)=&-f^{(2)}(t)\left[\left(\lambda^{(2)}\cos\hat\alpha\cos\hat\beta+\Lambda^{(2)}\sin\hat\alpha\sin\hat\beta\right)\ad\tau_-^2 + \left(\Lambda^{(1)}\cos\hat\alpha\cos\hat\beta+\lambda^{(1)}\sin\hat\alpha\sin\hat\beta\right)\ad\tau_+^2 \right.\notag\\
	&+\left.\left(\Lambda^{(2)}\sin\hat\alpha\cos\hat\beta-\lambda^{(2)}\cos\hat\alpha\sin\hat\beta\right)\ad\tau_+^1\tau_z^2+\left(\lambda^{(2)}\sin\hat\alpha\cos\hat\beta-\Lambda^{(2)}\cos\hat\alpha\sin\hat\beta\right)\ad\tau_-^1\tau_z^2+\mrm{H.c.}\right],
\end{align}
\begin{align}
	H_\mrm{PO}(t)=&-\left(a^2+\ad\,\!^2\right)\left\{\left[\lambda^{(1)}\Lambda^{(1)}f^{(1)}(t)\left(\cos2\hat\alpha+\cos2\hat\beta\right)-\lambda^{(2)}\Lambda^{(2)}f^{(2)}(t)\left(\cos2\hat\alpha-\cos2\hat\beta\right)\right]\frac{\tau_z^1}{2}\right.\notag\\
		&+\left[-\lambda^{(1)}\Lambda^{(1)}f^{(1)}(t)\left(\cos2\hat\alpha-\cos2\hat\beta\right)+\lambda^{(2)}\Lambda^{(2)}f^{(2)}(t)\left(\cos2\hat\alpha+\cos2\hat\beta\right)\right]\frac{\tau_z^2}{2}\notag\\
		&+2\sin\hat\alpha\cos\hat\alpha\left(\lambda^{(1)}\Lambda^{(1)}f^{(1)}(t)-\lambda^{(2)}\Lambda^{(2)}f^{(2)}(t)\right)\left(\tau_-^1\tau_+^2+\tau_+^1\tau_-^2\right)\notag\\
		&+\left.2\sin\hat\beta\cos\hat\beta\left(\lambda^{(1)}\Lambda^{(1)}f^{(1)}(t)+\lambda^{(2)}\Lambda^{(2)}f^{(2)}(t)\right)\left(\tau_-^1\tau_-^2+\tau_+^1\tau_+^2\right)\right\},
\end{align}
\begin{align}
	&H_\mrm{QQ}(t)=\left[-\frac{f^{(1)}(t)+f^{(2)}(t)}{2}x_0\cos2\hat\alpha	+\left(f^{(1)}(t)\hat s_n^{(1)}-f^{(2)}(t)\hat s_n^{(2)}\right)\sin\hat\alpha\cos\hat\alpha\right]\left(\tau_-^1\tau_+^2+\tau_+^1\tau_-^2\right)\notag\\
	&\qquad+\left[-\frac{f^{(1)}(t)+f^{(2)}(t)}{2}x_1\cos2\hat\beta 
	+ \left(f^{(1)}(t)\hat s_n^{(1)}+f^{(2)}(t)\hat s_n^{(2)}\right)\sin\hat\beta\cos\hat\beta\right]\left(\tau_-^1\tau_-^2+\tau_+^1\tau_+^2\right),
\end{align}
\begin{align}
	H_\mrm{QQ}^\phi(t)=&-\left(f^{(1)}(t)+f^{(2)}(t)\right)\left\{x_0\cos\hat\alpha\sin\hat\alpha\left[\left(\cos^2\hat\beta\tau_+^1\tau_-^1+\sin^2\hat\beta\tau_-^1\tau_+^1\right)\tau_z^2\right.-\tau_z^1\left(\cos^2\hat\beta\tau_+^2\tau_-^2+\sin^2\hat\beta\tau_-^2\tau_+^2\right)\right]\notag\\
	&\qquad-x_1\cos\hat\beta\sin\hat\beta \left[\left(\sin^2\hat\alpha\tau_+^1\tau_-^1+\cos^2\hat\alpha\tau_-^1\tau_+^1\right)\tau_z^2-\left.\tau_z^1\left(\cos^2\hat\alpha\tau_+^2\tau_-^2+\sin^2\hat\alpha\tau_-^2\tau_+^2\right)\right]\right\},
\end{align}
\end{widetext}
where $x_0=\lambda^{(1)}\lambda^{(2)}-\Lambda^{(1)}\Lambda^{(2)}$ and $x_1=\lambda^{(1)}\Lambda^{(2)}+\lambda^{(1)}\Lambda^{(2)}$. We notice the presence of an additional term $H^\phi_\mrm{QQ}(t)$ with respect to the result of Eq.~\eqref{eq:FC:H} which represents oscillating shifts on qubit $i$ conditional on the state of qubit $j$. However, in the limit $J\ll\Delta_Q,\Sigma_Q$, these terms are of fourth order in $g^{(1)}$ and $g^{(2)}$ and are thus negligible.

\section{Dispersive theory for two many-level systems coupled to a resonator beyond the RWA\label{sec:disp:mls}}

As explained in Section~\ref{sec:FC:MLS}, the two-level dispersive theory of Section~\ref{sec:gen:disp} can be extended to the case of two many-level systems coupled to a resonator. The Hamiltonian of this system, described in Eq.~\eq{eqn:H:MLS}, can be partly diagonalized by the following dispersive transformation
\begin{align}	\label{eq:transform}
	U_\mrm D^\mrm{MLS} &= \exp\left[\sum_{k=1,2}G_\mrm{R}^{(k)}+G_\mrm{CR}^{(k)}\right.\notag\\
		&\qquad+\left.G_{\xi r}^{(k)}+G_{\xi a}^{(k)}+G_\mrm{R2}^{(k)}+G_\mrm{CR2}^{(k)}\right].
\end{align}
The first two terms in the exponential are the familiar terms introduced in Ref.~\cite{PhysRevA.80.033846}, here extended to a MLS.
\begin{align}
	G_\mrm{R}^{(k)}&=\sum_{i\in\mathcal H_{M-1}}\lambda_i^{(k)}\Pi_{i,i+1}^{(k)}\ad-\left(\lambda_i^{(k)}\right)^\ast\Pi_{i+1,i}^{(k)} a\\
	G_\mrm{CR}^{(k)}&=\sum_{i\in\mathcal H_{M-1}}\Lambda_i^{(k)}\Pi_{i,i+1}^{(k)}a-\left(\Lambda_i^{(k)}\right)^\ast\Pi_{i+1,i}^{(k)} \ad,
\end{align}
with all dispersive parameters defined in Section~\ref{sec:FC:MLS}. Including only these terms would lead to incomplete diagonalization of the Hamiltonian to second order in the couplings. Indeed, we would obtain a resonator squeezing term~\cite{PhysRevA.80.033846} and terms involving jump operators between second-nearest-neighbour MLS states~\cite{koch2007charge}. To cancel these terms, we need to add second-order corrections to the dispersive transform. These are
{\allowdisplaybreaks
\begin{align}
	G_{\xi r}^{(k)}&=\sum_{i\in\mathcal H_{M}}\xi_j^{(k)} a^2\Pi_{i,i}^{(k)}-\mrm{H.c.},\\
	G_{\xi a}^{(k)}&=\sum_{i\in\mathcal H_{M-2}}\left(\xi_i'^{(k)}\ad a+\xi_i''^{(k)}\right)\Pi_{i,i+2}^{(k)}-\mrm{H.c.},\\
	G_\mrm{R2}^{(k)}&=\sum_{i\in\mathcal H_{M-2}}\zeta_i^{(k)}\Pi_{i,i+2}^{(k)}\ad\,\!^2-\mrm{H.c.},\\
	G_\mrm{CR2}^{(k)}&=\sum_{i\in\mathcal H_{M-2}}\zeta_i'^{(k)}\Pi_{i,i+2}^{(k)}a^2-\mrm{H.c.},
\end{align}}
where we have defined
\begin{align}
	\xi_i^{(k)}&=\frac{\chi_{i-1}^{(k)}+\mu_{i-1}^{(k)}-\chi_i^{(k)}-\mu_i^{(k)}}{4\omega_r},\\
	\xi_i'^{(k)}&=\frac{\eta_i^{(k)}+\eta_i'^{(k)}}{2\left(\omega_i^{(k)}-\omega_{i+2}^{(k)}\right)},\\
	\xi_i''^{(k)}&=\frac{g_i^{(k)}\lambda_{i+1}^{(k)}-g_{i+1}^{(k)}\Lambda_i^{(k)}}{2\left(\omega_i^{(k)}-\omega_{i+2}^{(k)}\right)},
\end{align}
\begin{align}
	\zeta^{(k)}&=\frac{\eta_i^{(k)}}{2\left(\omega_{i+2}^{(k)}-\omega_i^{(k)}-2\omega_r\right)},\\
	\zeta_i'^{(k)}&=\frac{\eta_i'^{(k)}}{2\left(\omega_{i+2}^{(k)}-\omega_i^{(k)}+2\omega_r\right)},
\end{align}
and
\begin{align}
	\eta_i^{(k)}&=g_i^{(k)}\lambda_{i+1}^{(k)}-g_{i+1}^{(k)}\lambda_i^{(k)},\\
	\eta_i'^{(k)}&=g_i^{(k)}\Lambda_{i+1}^{(k)}-g_{i+1}^{(k)}\Lambda_i^{(k)}.
\end{align}
We note that
\begin{align}
	\eta_i^{(k)}=\lambda_i^{(k)}\lambda_{i+1}^{(k)}[(\omega_{i+1}^{(k)}-\omega_i^{(k)})-(\omega_{i+2}^{(k)}-\omega_{i+1}^{(k)})],
\end{align}
and similarly for $\eta_i'$. This means that for a MLS that can be described as a weakly anharmonic resonator, such as the transmon, terms involving $\eta_i$ or $\eta_i'$ are negligible. More precisely, these terms can be safely neglected when $|(\omega_{i+2}-\omega_{i+1})-(\omega_{i+1}-\omega_{i})|\ll\omega_r,\omega_i\;\forall\;i$. 

Using the Campbell-Baker-Hausdorf formula, we find that, to second order in the couplings, the generalized dispersive Hamiltonian is
\begin{align}\label{eqn:MLS:coupled}
	&H_\mrm D =\omega_r\ad a+\sum_{k=1,2}\sum_{i\in\mathcal H_{M}}\left[\tilde{\omega}_i^{(k)}\Pi_{i,i}^{(k)}+S_i^{(k)}\Pi_{i,i}^{(k)}\ad a\right]\notag\\
			 &+ \sum_{i,j\in\mathcal H_{M-1}}J_{ij}\Pi_{i,i+1}^{(1)}\Pi_{j+1,j}^{(2)}+J'_{ij}\Pi_{i,i+1}^{(1)}\Pi_{j,j+1}^{(2)}+\mrm{H.c.},
\end{align}
where
\begin{align}	\label{eqn:transform:J}
	J_{ij}&=\frac12g_i^{(1)}\left(\lambda_j^{(2)}-\Lambda_j^{(2)}\right)^\ast\!\!\! + \frac12g_i^{(2)}\left(\lambda_j^{(1)}-\Lambda_j^{(1)}\right)^\ast\!\!,\\
	J'_{ij}&=\frac12g_i^{(1)}\left(\lambda_j^{(2)}-\Lambda_j^{(2)}\right) + \frac12g_i^{(2)}\left(\lambda_j^{(1)}-\Lambda_j^{(1)}\right).
\end{align}

This Hamiltonian is not yet diagonal because of the $J_{ij}$ and $J'_{ij}$ terms. To complete the diagonalization, we assume that the detunings between transitions in the first and the second MLS are all much larger than $J_{ij}$ and $J'_{ij}\;\forall\;i,j$. Then, we can apply the unitary transformation
\begin{align}
&	U_\mrm{J}^\mrm{MLS}=\\
&	\exp\left[\sum_{i,j\in\mathcal H_{M-1}}\lambda^J_{ij}\Pi_{i,i+1}^{(1)}\Pi_{j+1,j}^{(2)}+\Lambda^J_{ij}\Pi_{i,i+1}^{(1)}\Pi_{j,j+1}^{(2)} 
	-\mrm{H.c.}\right],\notag
\end{align}
where
\begin{align}
	\lambda^J_{ij}&=\frac{J_{ij}}{\tilde{\omega}^{(1)}_{i+1,i}-\tilde{\omega}^{(2)}_{i+1,i}}\\
	\Lambda^J_{ij}&=\frac{J'_{ij}}{\tilde{\omega}^{(1)}_{i+1,i}+\tilde{\omega}^{(2)}_{i+1,i}}.
\end{align}
To second order in the couplings, this transformation suppresses the interaction between the two MLS and brings Eq.~\eq{eqn:MLS:coupled} to its diagonal form Eq.~\eq{eqn:diag:MLS}.

\bibliography{paper_arXiv}

\end{document}